\documentclass[12pt]{article}
\usepackage{epsfig}
\def\etal{{\it et~al.~}}

\def\GeV{\,{\rm GeV}}

\def\MeV{\,{\rm MeV}}
\def\sec{\,{\rm sec}}
\def\Gyr{\,{\rm Gyr}}

\def\yrs{\,{\rm yrs}}
\def\rcm{\,{\rm cm}}
\def\km{\,{\rm km}}

\def\Mpc{\,{\rm Mpc}}

\def\eV{{\,\rm eV}}

\def\cmm2{{\,\rm cm^{-2}}}
\def\cm2{{\,{\rm cm}^2}}
\def\cmm3{{\,{\rm cm}^{-3}}}
\def\gcmm3{{\,{\rm g\,cm^{-3}}}}
\def\kms{\,{\rm km\,s^{-1}}}

\def\mpl{{m_{\rm Pl}}}

\def\la{\mathrel{\mathpalette\fun <}}
\def\ga{\mathrel{\mathpalette\fun >}}
\def\fun#1#2{\lower3.6pt\vbox{\baselineskip0pt\lineskip.9pt
  \ialign{$\mathsurround=0pt#1\hfil##\hfil$\crcr#2\crcr\sim\crcr}}}

\begin{document}
\pagestyle{empty}
\begin{center}
\bigskip

%\rightline{FERMILAB--Pub--98/***-A}
%\rightline{astro-ph/9901113}
%\rightline{to appear in {\it Reviews of Modern Physics (Centennial Volume)}}

\vspace{.2in}
{\Large \bf COSMOLOGY AT THE MILLENNIUM}
\bigskip

\vspace{.2in}
Michael S. Turner$^{1,2}$ and J. Anthony Tyson$^3$\\

\vspace{.2in}
{\it $^1$Departments of Astronomy \& Astrophysics and of Physics\\
Enrico Fermi Institute, The University of Chicago, Chicago, IL~~60637-1433}\\

\vspace{0.1in}
{\it $^2$NASA/Fermilab Astrophysics Center\\
Fermi National Accelerator Laboratory, Batavia, IL~~60510-0500}\\

\vspace{0.1in}
{\it $^3$Bell Labs, Lucent Technologies, Murray Hill, NJ~~07974  }\\

\end{center}

\vspace{.3in}
\centerline{\bf ABSTRACT}
\bigskip

One hundred years ago we did not know how stars generate energy, the
age of the Universe was thought to be only millions of years,
and our Milky Way galaxy was the only galaxy known.  Today,
we know that we live in an evolving and expanding Universe
comprising billions of galaxies, all held together by dark matter.
With the hot big-bang model, we can trace the evolution of
the Universe from the hot soup of quarks and leptons that existed a
fraction of a second after the beginning to the formation
of galaxies a few billion years later, and finally to the Universe we
see today 13 billion years after the big bang,
with its clusters of galaxies, superclusters,
voids, and great walls.
The attractive force of gravity acting on tiny primeval inhomogeneities
in the distribution of matter gave rise to all the
structure seen today.
A paradigm based upon deep connections between
cosmology and elementary particle physics -- inflation + cold
dark matter -- holds the promise of extending our understanding
to an even more fundamental level and much earlier times,
as well as shedding light on the unification of the forces
and particles of nature.  As we enter the 21st century, a flood of
observations is testing this paradigm.

\newpage
\pagestyle{plain}
\setcounter{page}{1}
\newpage

\centerline{DEDICATION}
\begin{quote}

This article is dedicated to the memory of a great cosmologist
and a very dear friend, David N. Schramm, who, had he not died
tragically in a plane crash, would have been a co-author of this
review.

\end{quote}
\vskip 1in

\section{INTRODUCTION}

One hundred years ago, we did not know how stars shine and
we had only a rudimentary understanding of one galaxy,
our own Milky Way.  Our knowledge of the Universe -- in
both space and time -- was scant:  Most of it
was as invisible as the world of the elementary particles.

Today, we know that we live in an evolving Universe filled
with billions of galaxies within our sphere of observation, and
we have recently identified the epoch when galaxies first appeared.
Cosmic structures from galaxies increasing in size to
the Universe itself are held together by invisible matter
whose presence is only known through its gravitational effects
(the so-called dark matter).  

The optical light we receive from the most distant galaxies takes us back to
within a few billion years of the beginning.  The microwave echo
of the big bang discovered by Penzias and Wilson in 1964
is a snapshot of the Universe at 300,000 years, long before
galaxies formed.  Finally, the light elements D, $^3$He, $^4$He and
$^7$Li were created by nuclear reactions even earlier and
are relics of the first seconds.  (The rest of the elements
in the periodic table were created in stars and stellar explosions
billions of years later.)

Crucial to the development of our understanding of the
cosmos, were advances in physics -- atomic, quantum, nuclear,
gravitational and elementary particle physics.
The hot big-bang model, based upon Einstein's
theory of General Relativity and supplemented by the aforementioned
microphysics, provides our quantitative
understanding of the evolution of the Universe from a
fraction a second after the beginning to the present, some
13 billion years later.  It is so successful that for more
than a decade it has been called the standard cosmology
(see e.g., Weinberg, 1972).

Beyond our current understanding, we are striving to answer
fundamental questions and test bold ideas based on the connections
between the inner space of the elementary particles
and the deep outer space of cosmology.  Is the ubiquitous dark
matter that holds the Universe together and determines its fate
composed of slowly moving elementary particles (called cold
dark matter) left over from
the earliest fiery moments?  Does all the structure seen in
the Universe today --- from galaxies to superclusters and great
walls --- originate from quantum mechanical fluctuations occurring during
a very early burst of expansion driven by vacuum
energy (called ''inflation'')?
Is the Universe spatially flat as predicted by inflation?
Does the absence of antimatter and the tiny ratio of
matter to radiation (around one part in $10^{10}$) involve forces
operating in the early Universe that violate baryon-number conservation
and matter -- antimatter symmetry?  Is inflation the dynamite
of the big bang, and if not, what is?  Is the expansion of the
Universe today accelerating rather than slowing, due to the presence
of vacuum energy or something even more mysterious?

Our ability to study the Universe has improved equally dramatically.
One hundred years ago our window on the cosmos consisted of visible
images take on photographic plates using telescopes of aperture
one meter or smaller.  Today, arrays of charge-coupled devices
have replaced photographic plates, improving photon collection
efficiency a hundredfold, and telescope apertures have grown
ten fold.  Together, they have increased photon collection by a factor
of $10^4$. Wavelength coverage has widened by a larger factor.  We now 
view the Universe with eyes that are sensitive from radio waves of length 
100 cm to gamma rays of energy up to $10^{12}\eV$,
from neutrinos to cosmic-ray particles; and perhaps someday 
via dark matter particles and gravitational radiation.

At all wavelengths advances in materials and device physics have spawned 
a new generation of low-noise, high-sensitivity detectors.
Our new eyes have opened new windows, allowing us to
see the Universe 300,000 years after the beginning,
to detect the presence of black holes, neutron stars and extra-solar planets, 
and to watch the birth of stars and galaxies.  One
hundred years ago the field of spectroscopy was in its infancy; today,
spectra of stars and galaxies far too faint even to be seen then, are
revealing the chemical composition and underlying physics of these
objects.  The advent of
computers and their dramatic evolution in power (quadrupling every
3 years since the 1970s) has made it possible to handle the data flow
from our new instruments as well as to analyze and to simulate the Universe.

This multitude of observations over the past decades has permitted 
cross-checks of our basic model of the Universe past as a denser, 
hotter environment
in which structure forms via gravitational instability driven by dark matter.
We stand on the firm foundation
of the standard big-bang model, with compelling ideas motivated by observations
and fundamental physics, as a flood of new observations looms.
This is a very exciting time to be a cosmologist.  Our late colleague
David N. Schramm more than once proclaimed the beginning of a golden age,
and we are inclined to agree with him.

\section{FOUNDATIONS}

There is now a substantial body of observations that support directly
and indirectly the relativistic hot Big Bang model for the expanding
Universe.  Equally important, there are no data that are inconsistent.
This is no mean feat:  The observations are sufficiently constraining
that there is no alternative to the hot Big Bang consistent
with all the data at hand.
Reports in the popular press of the death of the Big Bang usually
confuse detailed aspects of the theory that are still in a
state of flux, such as models
of dark matter or scenarios for large-scale structure formation, with
the basic framework itself.  There are indeed many open problems in
cosmology, such as the the age, size, and curvature of the Universe, 
the nature of the dark matter, and details of how large-scale 
structures form and how galaxies evolve --
these issues are being addressed by a number of current observations.
But the evidence that our Universe expanded from a dense hot
phase roughly 13 billion years ago is now incontrovertible
(see, e.g., Peebles \etal, 1991).

When studied with modern optical telescopes, the sky
is dominated by distant faint blue galaxies.
To 30th magnitude per square arcsecond surface brightness ($4 \times 10^{-18}$
erg sec$^{-1}$ cm$^{-2}$ arcsec$^{-2}$ in 100 nm bandwidth at 450 nm wavelength, 
or about five photons per minute per galaxy collected with a 4-meter mirror) 
there are about 50 billion galaxies over the sky.  
On scales less than around $100\Mpc$ galaxies
are not distributed uniformly, but rather cluster in a hierarchical
fashion.  The correlation length for bright galaxies is
$8h^{-1}\Mpc$ (at this distance from a galaxy the probability of
finding another galaxy is twice the average).  ($1\Mpc = 3.09\times
10^{24}\rcm \simeq 3$ million light years, and $h=H_0/100\kms\Mpc^{-1}$ is
the dimensionless Hubble constant.)

\begin{figure}
\epsfig{file=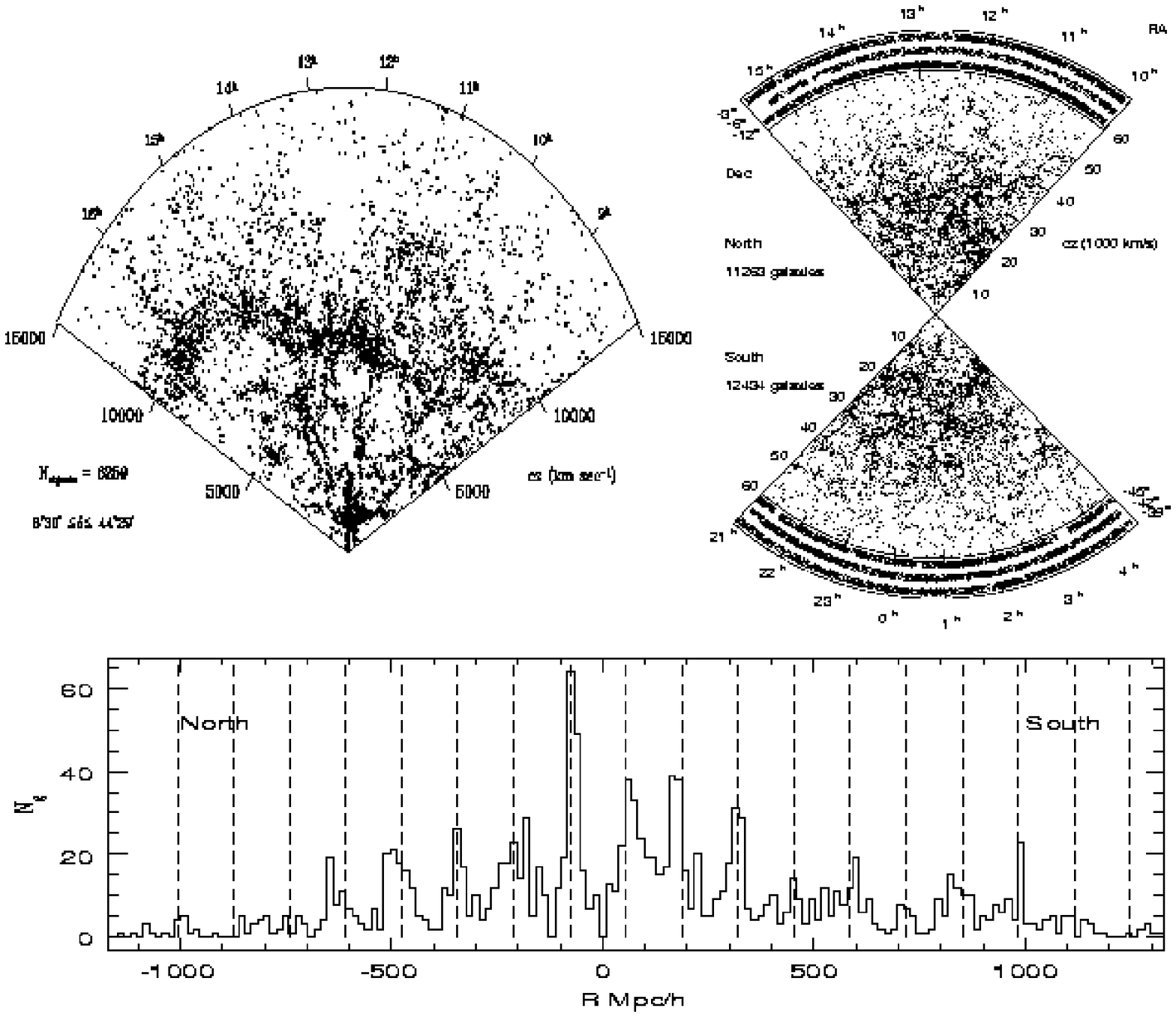,width=6.5in}
\caption{Large-scale structure in the Universe as traced by bright galaxies:
({\it upper left}) 
The Great Wall, identified by Geller and Huchra (1989) [updated by E. Falco].  
This coherent object stretches across most of the sky; walls of galaxies are the
largest known structures (see Oort, 1983).   
We are at the apex of the wedge, galaxies are placed at their `Hubble 
distances', $d=H_0^{-1}zc$; note too, the "voids" relatively devoid of
galaxies.  
({\it upper right}) 
Pie-diagram from the Las Campanas Redshift Survey (Shectman \etal, 1996).
Note the structure on smaller length scales including
voids and walls, which on larger scales dissolves into homogeneity.
({\it lower}) Redshift-histogram from deep pencil-beam surveys 
(Willmer, \etal, 1994; Broadhurst \etal, 1990) [updated by T. Broadhurst.]
Each pencil beam covers only a square degree on the sky.  The narrow width of
the beam `distorts' the view of the Universe, making it appear more
inhomogeneous.  The large spikes spaced by around $100h^{-1}\Mpc$
are believed to be great walls.
}
\label{fig:lss}
\end{figure}

About 10 percent of galaxies are found in clusters of galaxies, the largest 
of which contain thousands of galaxies.  Like galaxies, clusters are
gravitationally bound and no longer expanding.  Fritz Zwicky
was among the first to study clusters, and George Abell created the
first systematic catalogue of clusters of galaxies in 1958; since then, some
four thousand clusters have been identified (most discovered by
optical images, but a significant number by the x-rays emitted
by the hot intracluster gas).  Larger entities called superclusters,
are just now ceasing to expand and consist of several clusters.
Our own supercluster was first identified in 1937 by Holmberg, and 
characterized by de Vaucouleurs in 1953.  Other features in the distribution
of galaxies in 3-dimensional space have also been identified:
regions devoid of bright galaxies of size roughly 30 $h^{-1}$ $\Mpc$ (simply
called voids) and great walls of galaxies which stretch across a substantial
fraction of the sky and appear to be separated by about 100 $h^{-1}$ $\Mpc$.
Figure \ref{fig:lss} is a three panel summary of our knowledge of the
large-scale structure of the Universe.

\begin{figure}
\epsfig{file=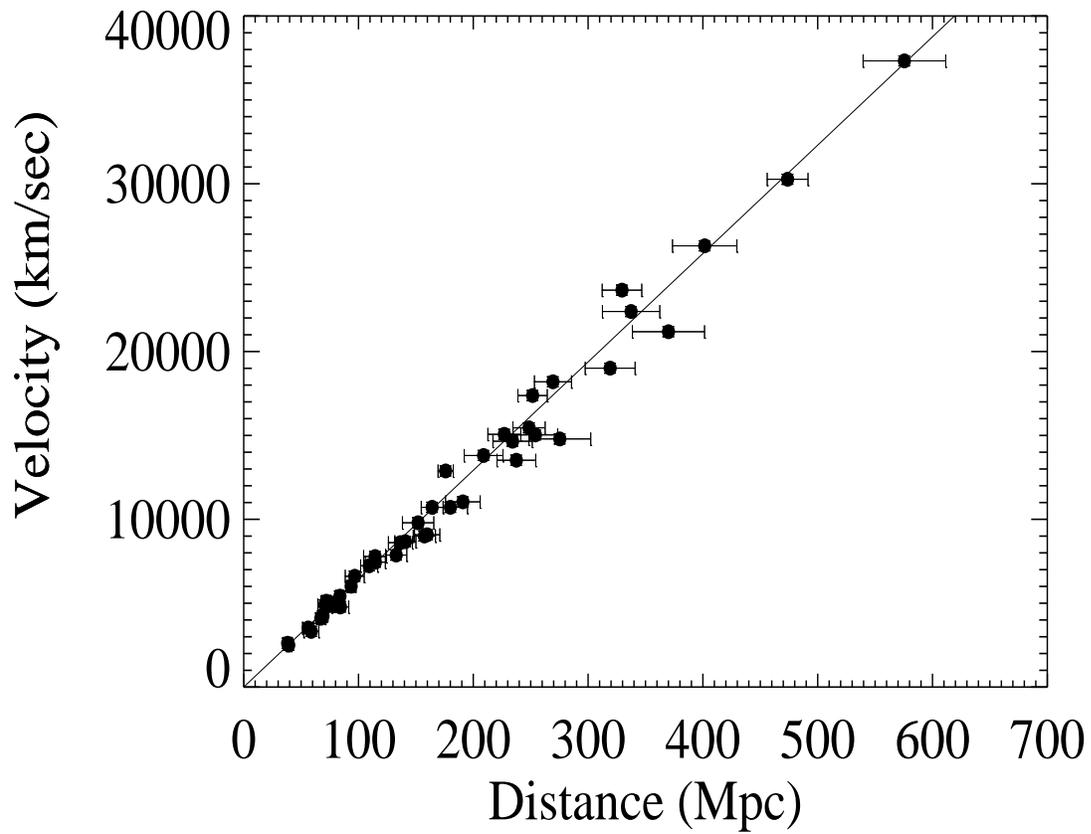,height=6in,width=5in,angle=90}
\caption{Hubble diagram based upon distances to supernovae of type 1a
(SNe1a).  Note the linearity; the slope, or Hubble constant,
$H_0 = 64\kms\Mpc^{-1}$ (Courtesy, A. Riess; see Filippenko \& Riess 1998).
}
\label{fig:H0}
\end{figure}

In the late 1920's Hubble established that
the spectra of galaxies at greater distances were systematically
shifted to longer wavelengths.  The change in wavelength of a spectral line
is expressed as the ``redshift'' of the observed feature,
\begin{equation}
1+z \equiv \lambda_{\rm observed}/ \lambda_{\rm emitted}.
\end{equation}
Interpreting the redshift as a Doppler velocity, Hubble's
relationship can be written
\begin{equation}
z \simeq H_0 d / c \qquad ({\rm for}\ z\ll 1).
\end{equation}
The factor $H_0$, now called the Hubble constant, is the expansion rate at
the present epoch.  Hubble's measurements of $H_0$ began
at $550\km\sec^{-1}\Mpc^{-1}$; a number of systematic errors
were identified, and by the 1960s $H_0$ had dropped to $100\kms\Mpc^{-1}$.
Over the last two decades controversy surrounded $H_0$, with
measurements clustered around $50\kms\Mpc^{-1}$ and $90\kms\Mpc^{-1}$.
In the past two years or so, much progress has been made because
of the calibration of standard candles by the Hubble Space Telescope
(see e.g., Filippenko and Riess, 1998; Madore \etal, 1998),
and there is now a general consensus that $H_0 = (67 \pm 10)\kms\Mpc^{-1}$
(where $\pm 10\kms\Mpc^{-1}$ includes both statistical and systematic
error; see Fig.~\ref{fig:H0}).
The inverse of the Hubble constant -- the Hubble time --
sets a timescale for the age of the Universe:  $H_0^{-1} = (15\pm 2) \Gyr$.

By now, through observations of a variety of phenomena from optical galaxies
to radio galaxies, the cosmological interpretation of redshift is very well
established.  Two recent interesting observations provide further
evidence:  numerous examples of high-redshift objects being
gravitationally lensed by low redshift objects near the line of sight;
and the fading of supernovae of type Ia, whose light curves
are powered by the radioactive decay of Ni$^{56}$, at high redshift
exhibiting time dilation by the predicted factor of $1+z$
(Leibundgut, \etal, 1996).

\begin{figure}
\epsfig{file=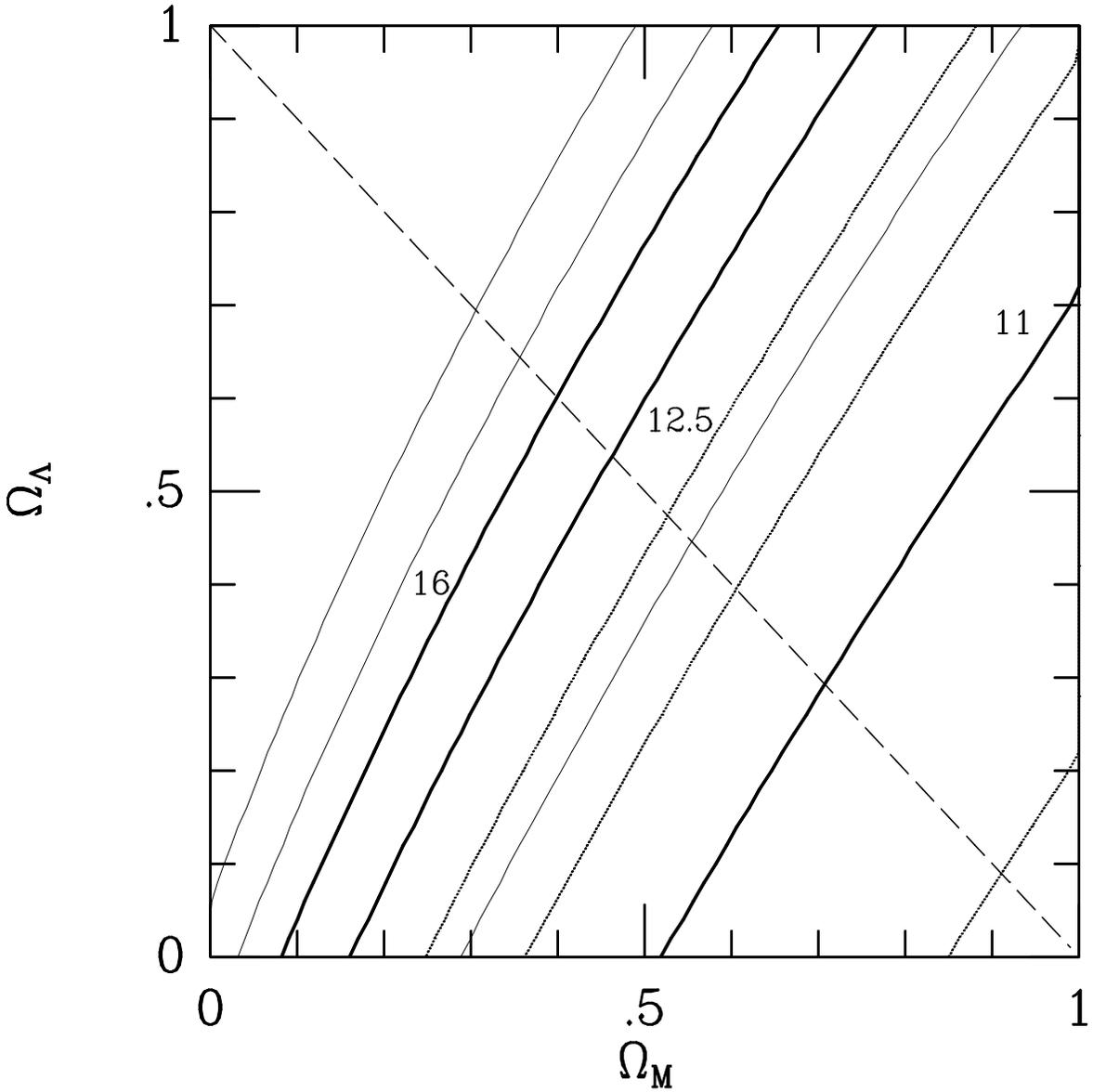,height=6in,width=6in}
\caption{
Contours of constant time back to the big bang in
the $\Omega_M$ -- $\Omega_\Lambda$ plane.  The three bold solid lines
are for $h=0.65$; the light solid lines are for $h=0.7$; and the dotted
lines are for $h=0.6$.
The diagonal line corresponds to a flat Universe.
Note, for $h\sim 0.65$ and $t_0\sim 13\Gyr$ a flat Universe is
possible only if $\Omega_\Lambda \sim 0.6$; $\Omega_M =1$ is
only possible if $t_0\sim 10\Gyr$ and $h\sim 0.6$.
}
\label{fig:age-H0}
\end{figure}

An important consistency test of the standard cosmology is the
congruence of the Hubble time with other independent determinations
of the age of the Universe.  (The product of the Hubble constant
and the time back to the big bang, $H_0 t_0$, is expected to be
between $2/3$ and $1$, depending upon the density of matter in the Universe;
see Fig.~\ref{fig:age-H0}.)  Since the discovery of the expansion, there have
been occasions when the product $H_0t_0$ far exceeded unity,
indicating an inconsistency.  Both $H_0$ and $t_0$ measurements have been 
plagued by systematic errors.  Slowly, the situation has
improved, and at present there is consistency within the
uncertainties.  
%Of the two, the $t_0$ measurement is currently the least accurate.
Chaboyer \etal (1998) date the oldest globular stars at $11.5 \pm 1.3 \Gyr$;
to obtain an estimate of the age of the Universe,
another $1-2\Gyr$ must be added to account for the time to
the formation of the oldest globular clusters.  Age estimates
based upon abundance ratios of radioactive isotopes produced
in stellar explosions, while dependent upon the time history of
heavy-element nucleosynthesis in our galaxy, provide a
lower limit to the age of the Galaxy of $10\Gyr$ (Cowan \etal, 1991).
Likewise, the age of the Galactic disk based upon the cooling
of white dwarfs, $> 9.5 \Gyr$, is also consistent with
the globular cluster age (Oswalt \etal, 1996).  Recent type Ia supernova
data yield an expansion age for the Universe of $14.0 \pm 1.5$ Gyr, including 
an estimate of systematic errors (Riess \etal, 1998).

Within the uncertainties, it is still possible that $H_0t_0$
is slightly greater than one.  This could either indicate a fundamental
inconsistency or the presence of a cosmological constant (or
something similar).  A cosmological constant can lead to
accelerated expansion and $H_0t_0 > 1$.  Recent measurements of
the deceleration of the Universe, based upon the distances of
high-redshift supernovae of type Ia (SNe1a), in fact show
evidence for accelerated expansion; we will return to these
interesting measurements later.

\begin{figure}
\epsfig{file=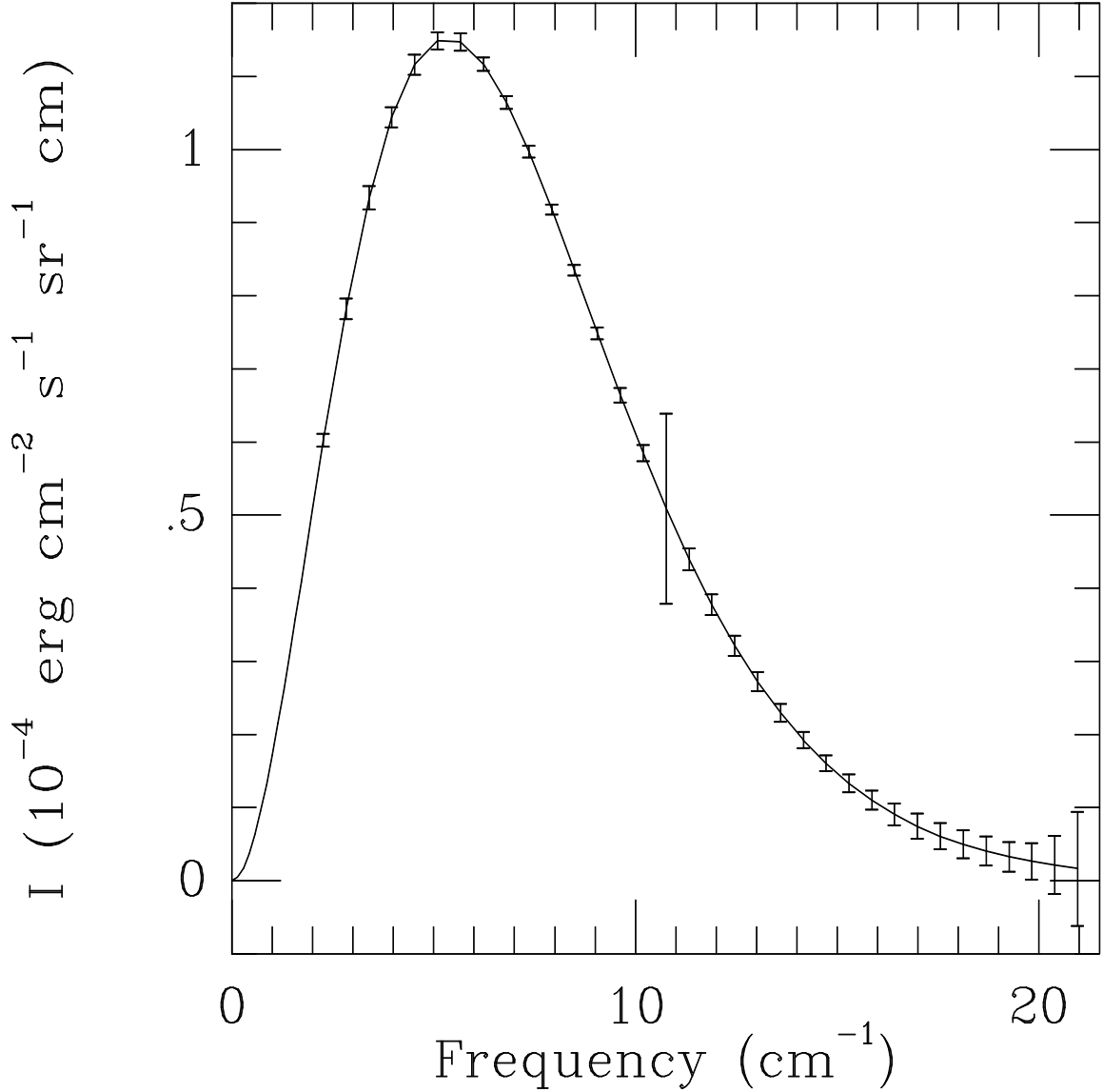,height=6in,width=6in,angle=-90}
\caption{Spectrum of the Cosmic Microwave Background Radiation
as measured by the FIRAS instrument on COBE and a black body
curve for $T=2.7277\,$K.  Note, the error flags have been enlarged
by a factor of 400.  Any distortions from the Planck curve are
less than 0.005\% (see Fixsen \etal, 1996).
}
\label{fig:CMB-spec}
\end{figure}

Another observational pillar of the Big Bang is the 2.73 K cosmic microwave
background radiation [CMB] (see Wilkinson, 1999).
The FIRAS instrument on the Cosmic Background Explorer [COBE] satellite has 
probed the CMB to extraordinary precision (Mather, \etal 1990).  
The observed CMB spectrum is 
exquisitely Planckian:  any deviations are smaller than 300 parts per 
million (Fixsen \etal, 1996), and the temperature is $2.7277 \pm 0.002\,$K
(see Fig.~\ref{fig:CMB-spec}).
The only viable explanation for such perfect black-body radiation is the
hot, dense conditions that are predicted to exist at early times
in the hot Big Bang model.  The CMB photons last scattered (with
free electrons)
when the Universe had cooled to a temperature of around $3000\,$K
(around 300,000 years after the Big Bang),
and ions and electrons combined to form neutral atoms.  Since
then the temperature decreased as $1+z$, with the expansion preserving
the black body spectrum.  The cosmological redshifting of the
CMB temperature was confirmed by a measurement of a temperature
of $7.4 \pm 0.8$ K at redshift 1.776 (Songaila \etal, 1994)
and of $7.9\pm 1$\,K at redshift 1.973 (Ge \etal, 1997),
based upon the population of hyperfine states in neutral carbon atoms
bathed by the CMB.

\begin{figure}
\epsfig{file=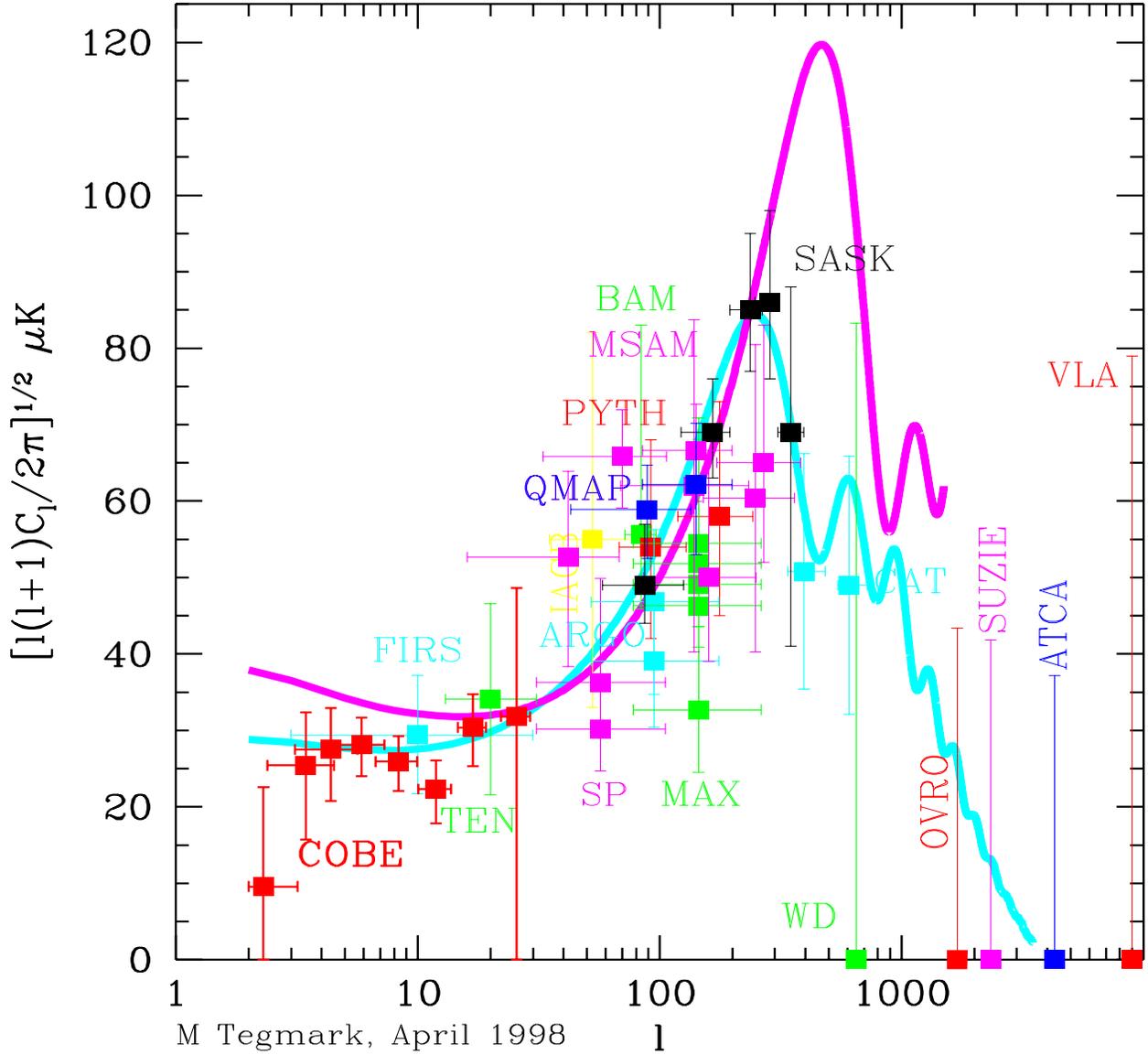,height=6in,width=6.5in}
\caption{Summary of current measurements of the power spectrum
of CMB temperature variations across the sky against spherical
harmonic number $l$ for several experiments.  The first acoustic peak is
evident.  The light curve, which is preferred by the data,
is a flat Universe ($\Omega_0 =1$, $\Omega_M = 0.35$), and the dark 
curve is for a open Universe ($\Omega_0=0.3$) [courtesy of M. Tegmark].
}
\label{fig:CMB-anis}
\end{figure}

The CMB is a snapshot of the Universe at 300,000 yrs.  From the
time of its discovery, its uniformity across the sky (isotropy)
was scrutinized.  The first anisotropy discovered was dipolar
with an amplitude of about $3\,$mK, whose simplest interpretation
is a velocity with respect to the cosmic rest frame.  The FIRAS
instrument on COBE has refined this measurement to high precision:
the barycenter of the solar system moves at a velocity of $370\pm
0.5\kms$. Taking into account our motion around the center of
the Galaxy, this translates to a motion of $620\pm 20\kms$ for
our local group of galaxies.  
After almost thirty years of searching, firm evidence for
primary anisotropy in the CMB, at the level of $30\mu$K 
(or $\delta T/T \simeq 10^{-5}$) on angular scales of $10^\circ$ 
was found by the DMR instrument on COBE (see Fig.~\ref{fig:CMB-anis}).
The importance of this discovery was
two-fold.  First, this is direct evidence that the Universe at early
times was extremely smooth since density variations manifest
themselves as temperature variations of the same magnitude.
Second, the implied variations in the density were of the correct
size to account for the structure that exists in the Universe today:
According to the standard cosmology the structure seen today grew
from small density inhomogeneities ($\delta \rho /\rho \sim 10^{-5}$)
amplified by the attractive action of gravity over the past $13\Gyr$.

\begin{figure}
\epsfig{file=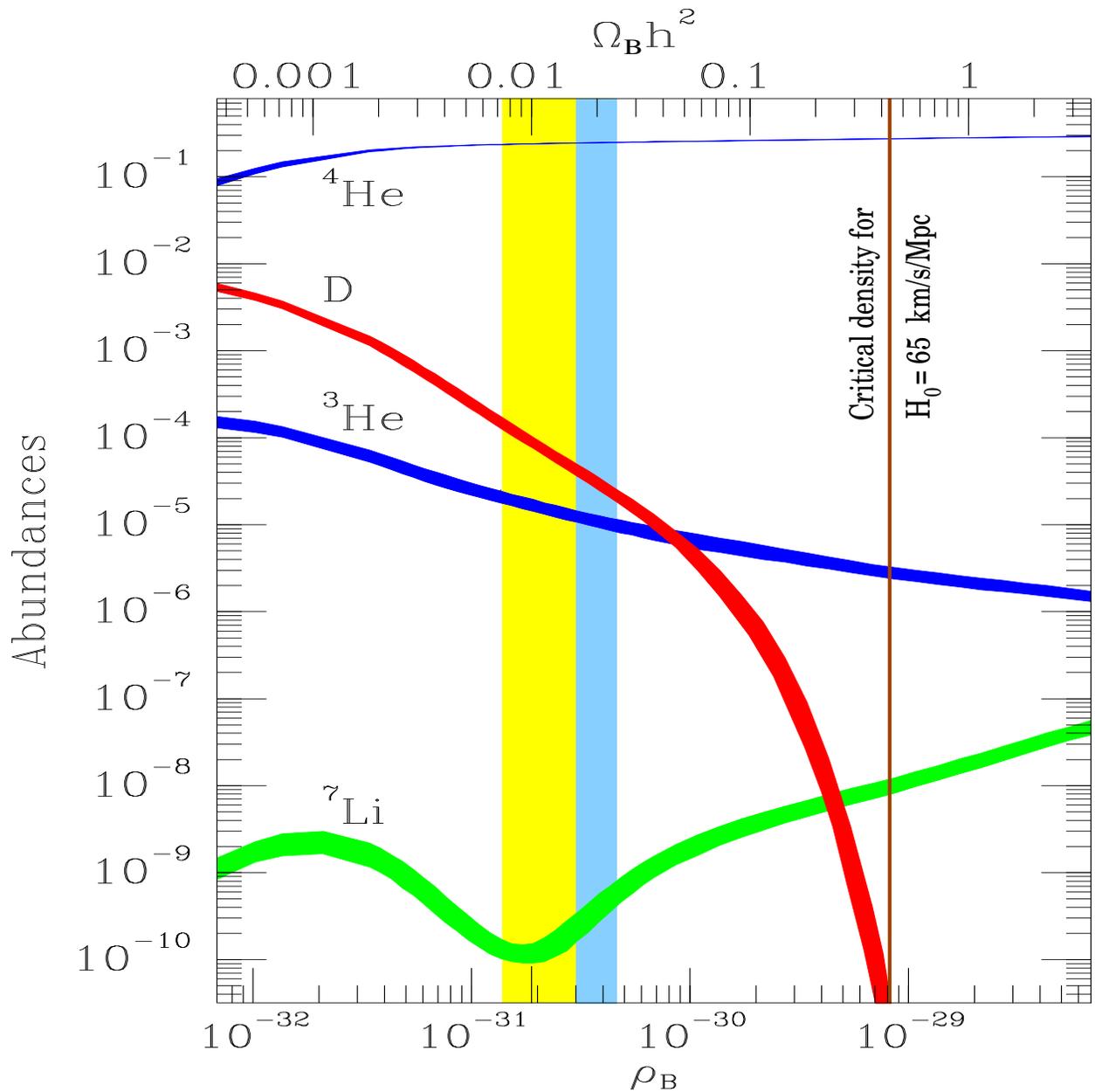,height=6.5in,width=6.5in}
\caption{Predicted abundances of $^4$He (mass fraction), D, $^3$He, and $^7$Li
(relative to hydrogen) as a function of the baryon density.
The broader band denotes the concordance interval based upon all
four light elements.  The narrower, darker band highlights
the determination of the baryon density based upon
a measurement of the primordial abundance of the most sensitive of
these  -- deuterium (Burles \& Tytler, 1998a,b), which implies
$\Omega_Bh^2 = 0.02\pm 0.002$.
}
\label{fig:BBN}
\end{figure}

The final current observational pillar of the standard cosmology is
big-bang nucleosynthesis [BBN].  When the Universe was seconds old
and the temperature was around $1\MeV$ a sequence of nuclear reactions
led to the production of the light elements
D, $^3$He, $^4$He and $^7$Li.
In the 1940s and early 1950s, Gamow and his collaborators suggested
that nuclear reactions in the early Universe could account for
the entire periodic table; as it turns out Coulomb barriers and the lack
of stable nuclei with mass 5 and 8 prevent further nucleosynthesis.
In any case, BBN is a powerful and very early test of the standard
cosmology:  the abundance pattern of the light elements predicted
by BBN (see Fig.~\ref{fig:BBN}) is consistent
with that seen in the most primitive samples of the cosmos.
The abundance of deuterium is very sensitive to the density of
baryons, and recent measurements of the deuterium abundance
in clouds of hydrogen at high redshift 
(Burles \& Tytler 1998a,b) have pinned down the baryon density to
a precision of 10\%.

\begin{figure}
\epsfig{file=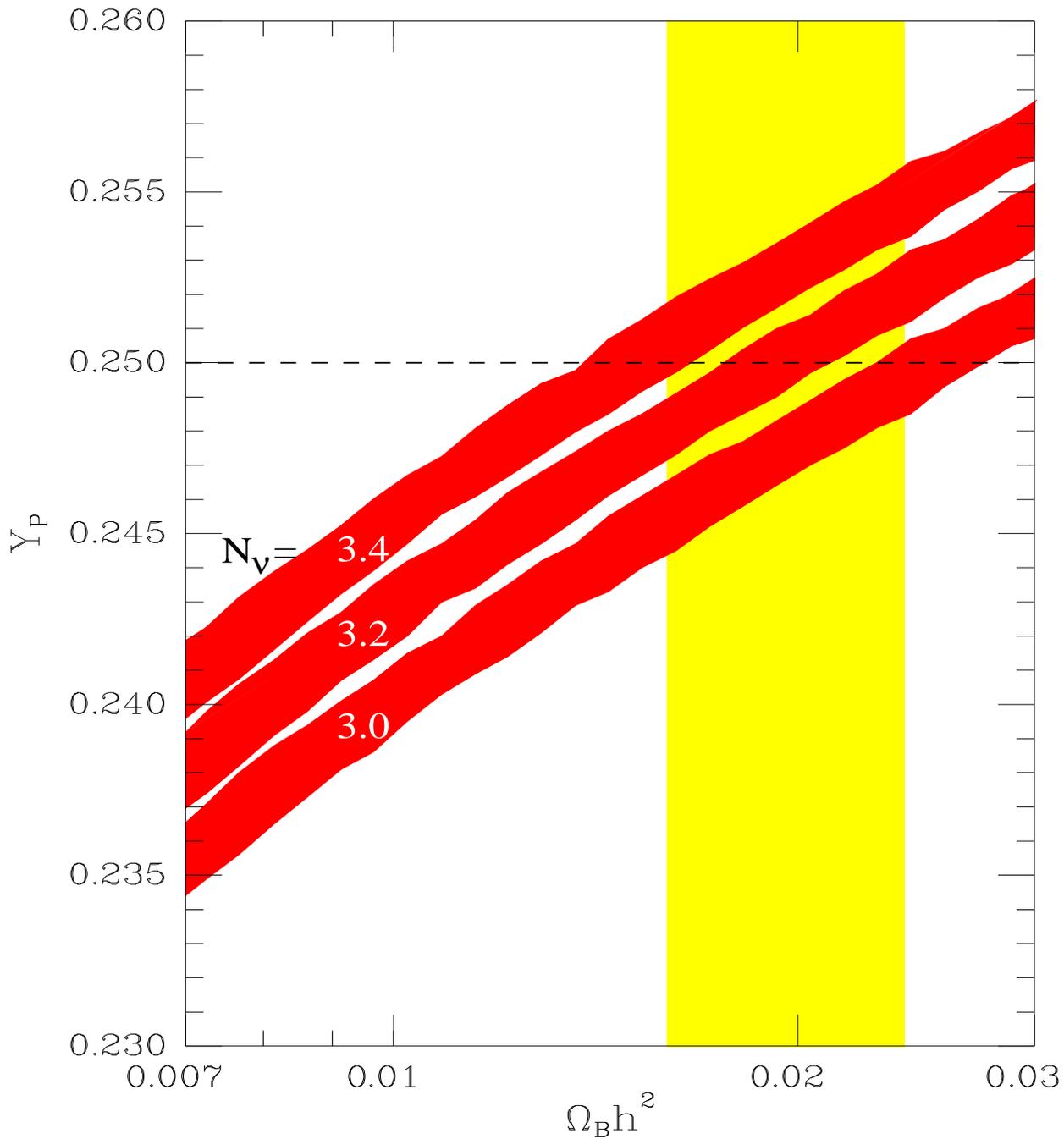,height=7in,width=6.5in}
\caption{The dependence of primordial $^4$He production, relative
to hydrogen, $Y_P$, on the number
of light neutrino species.  The vertical band denotes the
baryon density inferred from the Burles -- Tytler measurement
of the primordial deuterium abundance (Burles \& Tytler, 1998a,b);
using $Y_P<0.25$, based upon current $^4$He measurements,
the BBN limit stands at
$N_\nu < 3.4$ (from Schramm and Turner, 1998).
}
\label{fig:nulimit}
\end{figure}

As Schramm emphasized, BBN is also a powerful probe of fundamental
physics.   In 1977 he and his colleagues used BBN to place a
limit to the number of neutrino species (Steigman, Schramm and Gunn, 1977),
$N_\nu < 7$, which, at the time, was very poorly constrained by laboratory
experiments, $N_\nu$ less than a few thousand.  The limit is based upon the
fact that the big-bang $^4$He yield increases with $N_\nu$;
see Fig.~\ref{fig:nulimit}).  In 1989, experiments done
at $e^\pm$ colliders at CERN and SLAC
determined that $N_\nu$ was equal to three, confirming the cosmological
bound, which then stood at $N_\nu < 4$.  Schramm used the BBN limit
on $N_\nu$ to pique the interest of many particle physicists in
cosmology, both as a heavenly laboratory and in its own right.
This important cosmological constraint, and many others that followed,
helped to establish the ``inner space --
outer space connection'' that is now flourishing.

\section{THE STANDARD COSMOLOGY}

Most of our present understanding of the Universe is concisely
and beautifully summarized in the hot big-bang cosmological model
(see e.g., Weinberg, 1972; Peebles, 1993).
This mathematical description is based upon the
isotropic and homogeneous Friedmann-Lemaitre-Robertson-Walker
solution of Einstein's general relativity.  The evolution of
the Universe is embodied in the cosmic scale factor $R(t)$, which
describes the scaling up of all physical distances in the Universe
(separation of galaxies and wavelengths of photons).
The conformal stretching of the wavelengths of photons
accounts for the redshift of light from distant galaxies:
the wavelength of the radiation we see today is larger by
the factor $R(now) / R(then)$. Astronomers
denote this factor by $1+z$, which means that an object
at ``redshift z'' emitted the light seen today when the
Universe was a factor $1+z$ smaller.  Normalizing the
scale factor to unity today, $R_{\rm emission} = 1/(1+z)$.

It is interesting
to note that the assumption of isotropy and homogeneity was introduced
by Einstein and others to simplify the mathematics; as it turns out,
it is a remarkably accurate description at early times and today
averaged over sufficiently large distances (greater than 100 $\Mpc$ or so).

The evolution of the scale factor is governed by the Friedmann equation
for the expansion rate:
\begin{equation}
H^2 \equiv (\dot R / R)^2 = {8\pi G\rho \over 3} \pm {1\over R_{\rm curv}^2},
\label{F-eqn}
\end{equation}
where $\rho = \sum_i \rho_i$ is the total energy density from all
components of mass-energy, and $R_{\rm curv}$ is
the spatial curvature radius, which grows as the scale factor,
$R_{\rm curv} \propto R(t)$. [Hereafter we shall set c = 1.] As indicated
by the $\pm$ sign in Eq. (\ref{F-eqn})
there are actually three FLRW models; they differ in their spatial curvature:
The plus sign applies to the negatively curved model, and the minus
sign to the positively curved model.  For the spatially flat
model the curvature term is absent.

The energy density of a given component evolves according to
\begin{equation}
d\,\rho_i R^3 = -p_i d\, R^3  ,
\end{equation}
where $p_i$ is the pressure (e.g., $p_i \ll \rho_i$ for nonrelativistic
matter or $p_i = \rho_i/3$ for ultra-relativistic particles
and radiation).  The energy density of matter decreases as
$R^{-3}$, due to volume dilution.  The energy density of
radiation decreases more rapidly, as $R^{-4}$, the additional
factor arising because the energy of a relativistic particle
``redshifts'' with the expansion, $E\propto 1/R(t)$.  (This
of course is equivalent to the wavelength of a photon growing
as the scale factor.)  This redshifting of the energy
density of radiation by $R^{-4}$ also implies that
for black body radiation, the temperature decreases
as $T\propto R^{-1}$.

It is convenient to scale
energy densities to the critical density, $\rho_{\rm crit} \equiv
3H_0^2/8\pi G = 1.88h^2\times 10^{-29}\gcmm3\simeq 8.4\times
10^{-30}\gcmm3$ or approximately 5 protons per cubic meter,
\begin{eqnarray}
\Omega_i & \equiv & \rho_i /\rho_{\rm crit}\\
\Omega_0 & \equiv & \sum_i \Omega_i \\
R_{\rm curv} & = & H_0^{-1}/|\Omega_0 -1|^{1/2}
\end{eqnarray}
Note that the critical-density Universe ($\Omega_0 =1$) is flat;
the subcritical-density Universe ($\Omega_0 <1$) is negatively
curved; and the supercritical-density Universe ($\Omega_0 > 1$)
is positively curved.

There are at least two components to the energy density:
the photons in the $2.728\,$K cosmic microwave background radiation
(number density $n_\gamma = 412\cmm3$); and
ordinary matter in the formation of neutrons, protons and
associated electrons (referred to collectively as baryons).
The theory of big-bang nucleosynthesis and the
measured primordial abundance of deuterium imply that
the mass density contributed by baryons is $\Omega_B =
(0.02\pm 0.002)h^{-2} \simeq 0.05$.  In addition, the
weak interactions of neutrinos with electrons, positrons
and nucleons should have brought all three species of
neutrinos into thermal equilibrium when the Universe
was less than a second old, so that today there should
be three cosmic seas of relic neutrinos of comparable
abundance to the microwave photons, $n_\nu = {3\over 11}
n_\gamma \simeq 113\cmm3$ (per species).  (BBN provides a
nice check of this, because the yields depend sensitively upon the
abundance of neutrinos.)  Together, photons and neutrinos
(assuming all three species are massless, or very light, $\ll 10^{-3}\eV$)
contribute a very small energy density
$\Omega_{\nu \gamma} = 4.17h^{-2}\times 10^{-5} \simeq 10^{-4}$.

There is strong evidence for the existence of matter beyond
the baryons, as dynamical measurements of the matter density
indicate that it is at least 20\% of the critical density
($\Omega_M > 0.2$), which is far more than ordinary matter
can account for.  The leading explanation for the additional
matter is long-lived or stable elementary particles left over
from the earliest moments (see Section 5).

Finally, although it is now known that the mass density of the Universe
in the form of dark matter exceeds 0.2 of the closure density, there are 
even more exotic possibilities for additional components to the mass-energy
density, the simplest of which is Einstein's
cosmological constant.  Seeking static solutions, Einstein introduced
his infamous cosmological constant; after the discovery of the
expansion by Hubble he discarded it.  In the quantum world it
is no longer optional:  the cosmological constant represents
the energy density of the quantum vacuum (Weinberg, 1989; Carroll
\etal, 1992).  Lorentz invariance implies that the pressure associated
with vacuum energy is $p_{\rm VAC} = -\rho_{\rm VAC}$, and this ensures that
$\rho_{\rm VAC}$ remains constant as the Universe expands.  Einstein's
cosmological constant appears as an additional term $\Lambda /3$
on the right hand side of the Friedmann equation [Equ. 3]; it is equivalent
to a vacuum energy $\rho_{\rm VAC} = \Lambda /8\pi G$.
%%We know that this cosmological constant term cannot dominate the matter
%%term, since we live in a Universe with mass+gravity generated structure.

All attempts to calculate the
cosmological constant have been unsuccessful to say the very
least:  due to the zero-point energies the vacuum energy formally
diverges (``the ultraviolet catastrophe'').  Imposing a short wavelength
cutoff corresponding to the weak scale ($\sim 10^{-17}\rcm$)
is of little help:  $\Omega_{\rm VAC} \sim 10^{55}$!
The mystery of the cosmological constant is a fundamental one which
is being attacked from both ends:  Cosmologists are trying to
measure it, and particle physicists are trying to understand
why it is so small.

Because the different contributions to the energy density
scale differently with the cosmic scale factor, the expansion
of the Universe goes through qualitatively different phases.
While today radiation and relativistic particles are not
significant, at early times they dominated the energy, since
their energy density depends most strongly on the scale factor
($R^{-4}$ vs. $R^{-3}$ for matter).  Only at late times does
the curvature term ($\propto R^{-2}$) become important; for
a negatively curved Universe it becomes dominant.  For a
positively curved Universe, the expansion halts when it cancels
the matter density term and a contraction phase begins.

The presence of a cosmological constant, which is independent
of scale factor, changes this a little.  A flat or negatively
curved Universe ultimately enters an exponential expansion
phase driven by the cosmological constant.  This also occurs
for a positively curved Universe, provided the cosmological
constant is large enough,
\begin{equation}
\Omega_\Lambda > 4\Omega_M \left\{
\cos \left[ {1\over 3}\cos^{-1}(\Omega_M^{-1} -1)
+ {4\pi \over 3}\right] \right\}^3 .
\end{equation}
If it is smaller than this, recollapse
occurs.  Einstein's static Universe obtains for
$\rho_M=2\rho_{\rm VAC}$ and $R_{\rm curv} = 1/\sqrt{8\pi G\rho_{\rm VAC}}$.

The evolution of the Universe according to the standard hot big-bang
model is summarized as follows:

\begin{itemize}

\item {\em Radiation-dominated phase.}  At times earlier than
about $10,000\yrs$, when the temperature exceeded $k_BT \ga 3\eV$,
the energy density in radiation and relativistic particles exceeded
that in matter.  The scale factor grew as $t^{1/2}$ and the
temperature decreased as $k_BT \sim 1\MeV (t/\sec )^{-1/2}$.
At the earliest times, the energy in the Universe consists of
radiation and seas of relativistic particle -- antiparticle pairs.  (When
$k_BT \gg mc^2$ pair creation makes particle -- antiparticle
pairs as abundant as photons.)  The standard model of particle
physics, the $SU(3)\otimes SU(2) \otimes U(1)$ gauge theory
of the strong, weak and electromagnetic interactions, provides
the microphysics input needed to go back to $10^{-11}\sec$ when
$k_BT \sim 300\GeV$.  At this time the sea of relativistic
particles includes six species of quarks and antiquarks (up, down,
charm, strange, top and bottom), six types of leptons and antileptons
(electron, muon, and tauon and their corresponding neutrinos),
and twelve gauge bosons (photon, $W^\pm$, $Z^0$, and eight
gluons).
When the temperature drops below the mass of a particle species,
those particles and their antiparticles
annihilate and disappear (e.g., $W^\pm$ and $Z^0$ disappear
when $k_BT \sim mc^2 \sim 90\GeV$).  As the temperature
fell below $k_BT \sim 200\MeV$, a phase transition occurred from a 
quark-gluon plasma to neutrons, protons and pions, along with the leptons,
antileptons and photons.  At a temperature of $k_BT \sim 100\MeV$,
the muons and antimuons disappeared.  When the temperature was
around $1\MeV$ a sequence of events and nuclear reactions
began that ultimately resulted in the synthesis of D, $^3$He,
$^4$He and $^7$Li.  During BBN, the last of the particle -- antiparticle
pairs, the electrons and positrons, annihilated.

\item {\em Matter-dominated phase.}  When the temperature reached
around $k_BT \sim 3\eV$, at a time of around $10,000\,years$, the
energy density in matter began to exceed that in radiation.  At
this time the Universe was about $10^{-4}$ of its present size
and the cosmic-scale factor began to grow as $R(t)\propto t^{2/3}$.
Once the Universe became matter-dominated, primeval inhomogeneities in
the density of matter (mostly dark matter), shown to be of size 
around $\delta\rho
/\rho \sim 10^{-5}$ by COBE and other anisotropy experiments,
began to grow under the attractive influence of gravity
($\delta\rho /\rho \propto R$).  After 13 billion or so years
of gravitational amplification, these tiny primeval density
inhomogeneities developed into all the structure that we see
in the Universe today, galaxies, clusters of the galaxies,
superclusters, great walls, and voids.  Shortly after matter
domination begins, at a redshift $1+z \simeq 1100$, photons
in the Universe undergo their last-scattering off free electrons;
last-scattering is precipitated by the recombination of electrons
and ions (mainly free protons), which occurs at a
temperature of $k_BT\sim 0.3\eV$ because neutral atoms  are
energetically favored.  Before last-scattering, matter and radiation
are tightly coupled;
after last-scattering, matter and radiation are essentially decoupled.

\item {\em Curvature-dominated or cosmological constant
dominated phase.}  If the Universe is negatively curved and there
is no cosmological constant, then when the size of Universe is
$\Omega_M/(1-\Omega_M) \sim \Omega_M$ times its present size the epoch
of curvature domination begins (i.e., $R_{\rm curv}^{-2}$
becomes the dominant term on the right hand side
of Friedmann equation).  From this
point forward the expansion no longer slows and $R(t) \propto t$
(free expansion).  In the case of a cosmological constant
and a flat Universe, the cosmological constant becomes
dominant when the size of the Universe is $[\Omega_M/(1-\Omega_M)]^{1/3}$.
Thereafter, the scale factor grows exponentially.
In either case, further growth of density inhomogeneities that
are still linear ($\delta \rho /\rho < 1$) ceases.  The
structure that exists in the Universe is frozen in.

\end{itemize}

Finally, a comment on the expansion rate and the size of
the `observable Universe.'
The inverse of the expansion rate has units of time. The Hubble time, $H^{-1}$,
corresponds to the time it takes for the scale factor to roughly
double.  For a matter-, radiation-, or curvature-dominated
Universe, the age of the Universe (time back to zero scale factor)
is:  ${2\over 3}H^{-1}$, ${1\over 2}H^{-1}$, and $H^{-1}$ respectively.
The Hubble time also sets the size of the observable (or causally
connected) Universe:  the distance to the `horizon,'
which is equal to the distance that light could have traveled
since time zero, is $2t = H^{-1}$ for a radiation-dominated
Universe and $3t = 2H^{-1}$ for a matter-dominated Universe.
Paradoxically, although the size of the Universe goes to zero
as one goes back to time zero, the expansion rate is larger,
and so points separated by distance $t$ are moving apart faster
than light can catch up with them.

\section{INNER SPACE AND OUTER SPACE}

The ``hot'' in the hot big-bang cosmology makes
fundamental physics an inseparable part of the
standard cosmology.  The time -- temperature
relation, $k_BT \sim 1\MeV (t/\sec )^{-1/2}$,
implies that the physics of higher energies and shorter
times is required to understand the Universe
at earlier times:  atomic physics at $t\sim
10^{13}\sec$, nuclear physics at $t\sim 1\sec$,
and elementary-particle physics at $t< 10^{-5}\sec$.
The standard cosmology model itself is based upon Einstein's
general relativity, which embodies our deepest
and most accurate understanding of gravity.

The standard model of particle physics, which
is a mathematical description of
the strong, weak and electromagnetic
interactions based upon the $SU(3)\otimes SU(2)\otimes
U(1)$ gauge theory, accounts for all known
physics up to energies of about $300\GeV$ (Gaillard, Grannis
and Sciulli, 1999).  It provides the input microphysics for the standard
cosmology necessary to discuss events as early
as $10^{-11}\sec$.  It also provides a firm foundation
for speculations about the Universe at even earlier times.

A key feature of the standard model of particle
physics is asymptotic freedom: at high energies
and short distances, the interactions between the
fundamental constituents of matter -- quarks and
leptons -- are perturbatively weak.
This justifies approximating the early Universe
as hot gas of noninteracting particles (dilute
gas approximation) and opens the door to sensibly
speculating about times as early as $10^{-43}\sec$,
when the framework of general relativity becomes
suspect, since quantum corrections to this classical
description are expected to become important.

The importance of asymptotic freedom for early-Universe
cosmology cannot be overstated.  A little more than
twenty-five years ago, before the advent of quarks and leptons
and asymptotic freedom,
cosmology hit a brick wall at $10^{-5}\sec$ because
extrapolation to early times was nonsensical.  The problem
was twofold:  the finite size of nucleons and related
particles and the exponential rise in the number of `elementary
particles' with mass.  At around $10^{-5}\sec$, nucleons would
be overlapping, and with no understanding of the strong forces
between them, together with the the exponentially rising spectrum
of particles, thermodynamics became ill-defined at higher
temperatures.

The standard model of particle physics has provided particle
physicists with a reasonable foundation for speculating about
physics at even shorter distances and higher energies.  Their
speculations have significant cosmological implications, and --
conversely -- cosmology holds the promise to test some of their
speculations.  The most promising particle physics ideas
(see e.g., Schwarz \& Seiberg, 1999) and their cosmological
implications are:

\begin{itemize}

\item {\em Spontaneous Symmetry Breaking (SSB).}  A key idea, which is
not fully tested, is that most of the underlying symmetry in a
theory can be hidden because the vacuum state does not respect
the full symmetry; this is known as spontaneous symmetry breaking
and accounts for the carriers of the weak force, the $W^\pm$ and $Z^0$
bosons, being very massive.  (Spontaneous symmetry breaking is
seen in many systems, e.g., a ferromagnet at low temperatures:
it is energetically favorable for the spins to align thereby
breaking rotational symmetry.)
In analogy to symmetry breaking in a ferromagnet,
spontaneously broken symmetries
are restored at high temperatures.  Thus, it is likely that
the Universe underwent a phase transition at around $10^{-11}\sec$
when the symmetry of the electroweak theory was broken, $SU(2)\otimes
U(1) \rightarrow U(1)$.

\item {\em Grand unification.}  It is possible to unify the strong,
weak, and electromagnetic interactions by a larger gauge group, e.g.,
$SU(5), SO(10),$ or $E8$.  The advantages are twofold:
the three forces are described
as different aspects of a more fundamental force with a single
coupling constant, and the quarks and leptons are unified as they
are placed in the same particle multiplets.  If true, this would
imply another stage of spontaneous symmetry breaking, $G \rightarrow
SU(3)\otimes SU(2)\otimes U(1)$.  In addition, grand unified theories
(or GUTs) predict that baryon and lepton number are violated -- so
that the proton is unstable and neutrinos have mass --
and that stable topological defects
associated with SSB may exist, e.g., point-like defects
called magnetic monopoles, one-dimensional defects referred to
as ``cosmic'' strings, and and two-dimensional defects called
domain walls.  The cosmological implications of GUTs are manifold:
neutrinos as a dark matter component, baryon and lepton number violation
explaining the matter -- antimatter asymmetry of the Universe,
and SSB phase transitions producing topological defects that
seed structure formation or a burst of tremendous expansion
called inflation.
                 
\item{\em Supersymmetry.}  In an attempt to put bosons and fermions
on the same footing, as well as to better understand the
`hierarchy problem,' namely, the large gap between
the weak scale ($300\GeV$) and the Planck scale ($10^{19}\GeV$),
particle theorists have postulated supersymmetry, the symmetry
between fermions and bosons.  (Supersymmetry also appears to
have a role to play in understanding gravity.)  Since the
fundamental particles of the standard model of particle physics
cannot be classified as fermion -- boson pairs,
if correct, supersymmetry
implies the existence of a superpartner for every known particle,
with a typical mass of order $300\GeV$.
The lightest of these superpartners, is usually stable
and called `the neutralino.'
The neutralino is an ideal dark matter candidate.

\item{\em Superstrings, supergravity, and M-theory.}  The unification
of gravity with the other forces of nature has long been the holy
grail of theorists.  Over the past two decades there have been
some significant advances:  supergravity, an 11-dimensional
version of general relativity with supersymmetry, which unifies
gravity with the other forces; superstrings,
a ten-dimensional theory of relativistic strings, which unifies
gravity with the other forces in a self-consistent, finite theory;
and M-theory, an ill-understood, ``larger'' theory that encompasses
both superstring theory and supergravity theory.  An obvious
cosmological implication is the existence of additional spatial
dimensions, which today must be ``curled up'' to escape notice,
as well as the possibility of sensibly describing cosmology at times
earlier than the Planck time.

\end{itemize}

Advances in fundamental physics have been crucial to advancing
cosmology: e.g., general relativity led to the first self-consistent
cosmological models; from nuclear physics came big-bang nucleosynthesis;
and so on.  The connection between fundamental physics and cosmology
seems even stronger today and makes realistic the hope that much
more of the evolution of the Universe will be explained by fundamental
theory, rather than ad hoc theory that dominated cosmology before
the 1980s.  Indeed, the most promising paradigm for extending
the standard cosmology, inflation + cold dark matter, is deeply
rooted in elementary particle physics.

\section{DARK MATTER AND STRUCTURE FORMATION}

As successful as the standard cosmology is, it leaves important questions
about the origin and evolution of the Universe unanswered.  To an optimist,
these questions suggest that there is a grander
cosmological theory, which encompasses the hot big-bang model
and resolves these questions.
It can easily be argued that the most pressing issues
in cosmology are:  the quantity
and composition of energy and matter in the Universe, 
and the origin and nature of the density perturbations that
seeded all the structure in the Universe.  Cosmology is
poised for major progress on these two questions. 
Answering these questions will provide a window to see beyond
the standard cosmology.

\subsection{Dark matter and dark energy}

Our knowledge of the mass and energy content of the Universe
is still poor, but is improving rapidly (see Sadoulet, 1999).
We can confidently say that most of the matter in the Universe is
of unknown form and dark (see e.g., Dekel, Burstein and White,
1997; Bahcall \etal, 1993):  Stars (and closely related material)
contribute a tiny fraction of the critical density, $\Omega_{\rm lum}
=(0.003\pm 0.001) h^{-1}\simeq 0.004$, while the amount of matter known to
be present from its gravitational effects contributes around
ten times this amount, $\Omega_M = 0.35\pm 0.07$ (this error
flag is ours; it is meant to indicate 95\% certainty that
$\Omega_M$ is between 0.2 and 0.5).  The gravity
of dark matter is needed to hold together just about everything
in the Universe -- galaxies, clusters of galaxies, superclusters
and the Universe itself.  A variety of methods for determining
the amount of matter all seem to converge on $\Omega_M \sim 1/3$;
they include measurements of the masses of clusters of
galaxies and the peculiar motions of galaxies.  Finally, the
theory of big-bang nucleosynthesis and the recently measured
primeval abundance of deuterium pin down the baryon density
very precisely:  $\Omega_B = (0.02\pm 0.002)h^{-2} \simeq 0.05$.
The discrepancy between this number and dynamical measurements
of the matter density is evidence for nonbaryonic dark matter.

Particle physics suggests three dark matter candidates
(Sadoulet, 1999):
a $10^{-5}\eV$ axion (Rosenberg, 1998); a $10\GeV - 500\GeV$ neutralino
(Jungman, Kamionkowski, and Griest, 1996); and
a $30\eV$ neutrino.  These three possibilities are highly
motivated in two important senses:  first, the axion and neutralino
are predictions of fundamental theories that attempt to go
beyond the standard model of particle physics, as are neutrino
masses; and second, the relic abundances of the axion and neutralino
turn out to be within a factor of ten of the critical density  -- and similarly
for the neutrino, GUTs predict masses in the eV range, which
is what is required to make neutrinos a significant contributor
to the mass density.

\begin{figure}
\epsfig{file=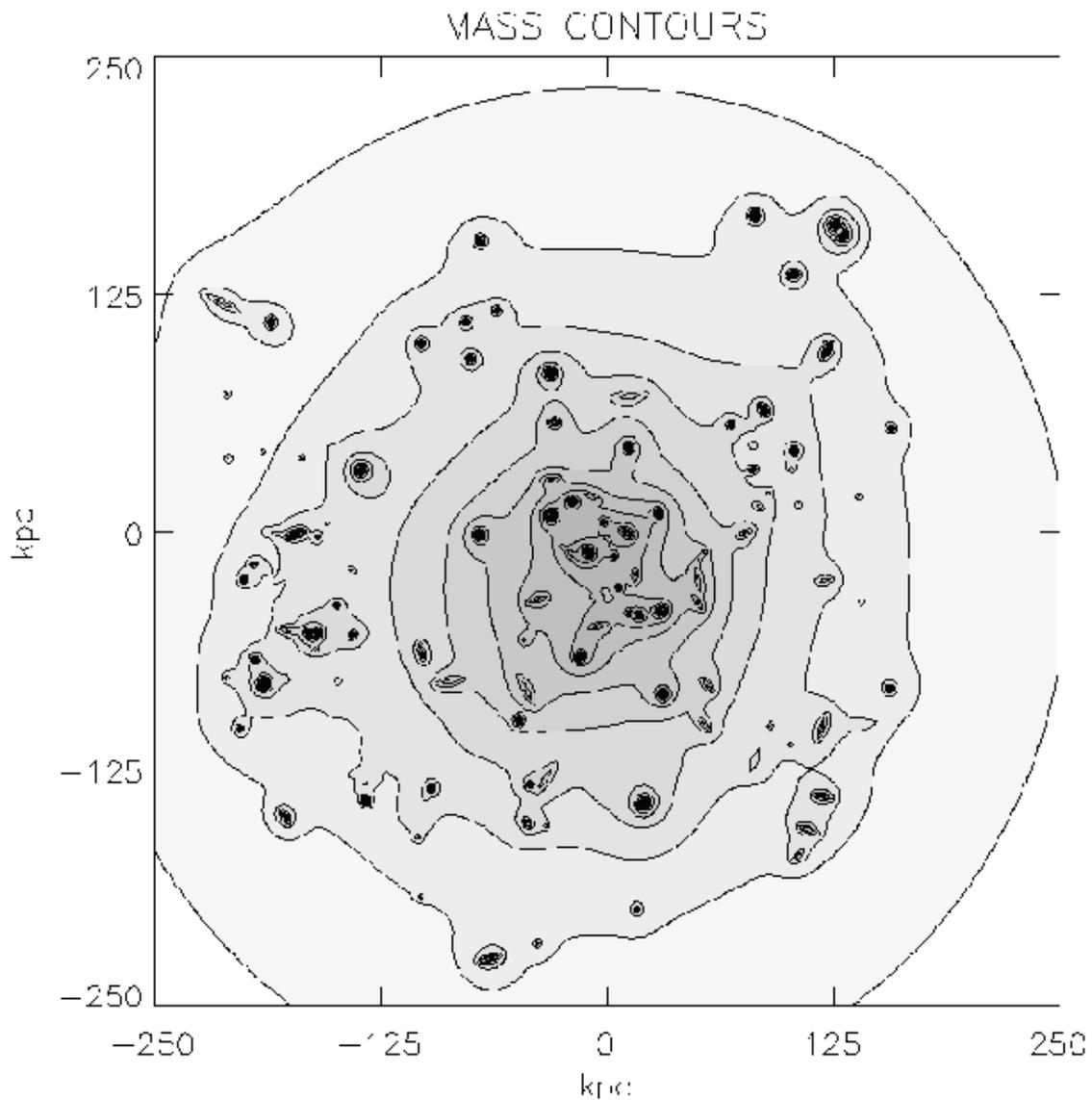,height=6in,width=6.5in}
\caption{
The reconstructed total mass density in the cluster of galaxies 0024+1654
at redshift z = 0.39, based on parametric inversion  of the associated 
gravitational lens.
%Mass is calibrated in units of the critical lensing density $\Sigma_C$ =
%2910 $M_\odot$ pc$^{-2}$.
Projected mass contours are spaced by 430 $M_\odot$ pc$^{-2}$, with the
outer contour at 1460 $M_\odot$ pc$^{-2}$.
Excluding dark mass concentrations centered on visible galaxies,
more than 98\% of the remaining mass is represented by a smooth
concentration of dark matter centered near the brightest cluster galaxies,
with a 50 kpc soft core (Tyson, \etal. 1998).   
}
\label{fig:0024}
\end{figure}

Because measuring the masses of galaxy clusters has been key to defining
the dark matter problems it is perhaps worth further discussion.
Cluster masses can be estimated by three
different techniques -- which give consistent results.  The first,
which dates back to Fritz Zwicky (1935), uses the measured velocities
of cluster galaxies and the virial theorem to determine the total
mass (i.e., $KE_{\rm gal}\simeq |PE_{\rm gal}|/2$).
The second method uses the temperature of the hot x-ray
emitting intracluster gas and the virial theorem to arrive at the total
mass.  The third and most direct method is using the gravitational
lensing effects of the cluster on much more distant galaxies.
Close to the cluster center, lensing is strong enough to produce
multiple images; farther out, lensing distorts the shape of distant
galaxies.  The lensing method allows the cluster (surface) mass
density to be mapped directly.  An example of mapping the mass
distribution of a cluster of galaxies is shown in Fig.~\ref{fig:0024}.

Using clusters to estimate the mean mass density of the Universe
requires a further assumption:  that their mass-to-light ratio provides
a good estimate for the mean mass-to-light ratio.  This is
because the mean mass density is determined by multiplying
the mean luminosity density (which is reasonably well measured)
by the inferred cluster mass-to-light ratio. Using this technique,
Carlberg \etal (1996, 1997) find $\Omega_M = 0.19 \pm 0.06 \pm 0.04$.
If clusters have more luminosity per mass than average, this technique
would underestimate $\Omega_M$.  

There is another way to estimate $\Omega_M$ using clusters, based on
a different, more physically motivated assumption.
X-ray measurements more easily determine the amount of hot, intracluster 
gas; and as it turns out, most the baryonic mass in a cluster resides
here rather than in the mass of individual galaxies (this fact is
also confirmed by lensing measurements).   Together with the
total cluster mass, the ratio of baryonic mass to total mass
can be determined; a compilation of the existing data give
$M_B/M_{\rm TOT} = (0.07\pm 0.007)h^{-3/2}\simeq 0.15$ (Evrard, 1997
and references therein).
Assuming that clusters provide a fair sample of matter in the Universe so
that $\Omega_B /\Omega_M = M_B/M_{\rm TOT}$, the accurate
BBN determination of $\Omega_B$ can be used to infer:  $\Omega_M
= (0.3\pm 0.05)h^{-1/2}\simeq 0.4 $.  [A similar result for
the cluster gas to total mass ratio is derived from cluster
gas measurements based upon the distortion of the CMB spectrum
due to CMB photons scattering off the hot cluster gas (Sunyaev -- Zel'dovich
effect); see Carlstrom, 1999.]

Two other measurements bear on the quantity and composition of
energy and matter in the Universe.  First, the pattern of anisotropy
in the CMB depends upon the total energy density in the
Universe (i.e., $\Omega_0$) (see e.g., Jungman, Kamionkowski,
Kosowsky and Spergel, 1996).  The peak in the multipole
power spectrum is $l_{\rm peak} \simeq 200/\sqrt{\Omega_0}$.
The current data, shown in Fig.~\ref{fig:CMB-anis},
are consistent with $\Omega_0
\simeq 1$, though $\Omega_0 \sim 0.3$ cannot be excluded.  This
together with the evidence that $\Omega_M\simeq 0.3$ 
leaves room for a component of energy that does not clump, such
as a cosmological constant.

The oldest approach to determining the total mass-energy density is 
through the deceleration parameter (Baum 1957; Sandage 1961),
which quantifies the present slowing of the expansion due to gravity,
\begin{equation}
q_0 \equiv -{(\ddot R / R)_0 \over H_0^2} = {\Omega_0 \over 2}
[1+3p_0/\rho_0]
\end{equation}
where subscript zero refers to quantities measured at the current epoch.
Note, in a Universe where the bulk of the matter is nonrelativistic
($p\ll \rho$), $q_0$ and $\Omega_0$ differ only by a factor of two.
The luminosity distance to an object at redshift $z\ll 1$ is related
to $q_0$,
\begin{equation}
d_L H_0 = z + z^2(1 - q_0)/2 + \cdots
\end{equation}
and thus accurate distance measurements can be used to determine $q_0$.
(The luminosity distance to an object is defined as that inferred from
the inverse square law:  $d_L \equiv \sqrt{{\cal L}/4\pi {\cal F}}$.)

Recently, two groups (The Supernova 
Cosmology Project and The High-z Supernova Team) using Type Ia supernovae
(SNe1a) as standard candles (objects of known ${\cal L}$) and
assuming that their flux measurements (i.e., ${\cal F}$)
were not contaminated by sample selection, evolution, or dust systematics,
both conclude that the expansion of the Universe
is accelerating rather than decelerating (i.e., $q_0 < 0$)
(Perlmutter \etal, 1998; Schmidt \etal, 1998).  If correct,
this implies that much of the energy in the Universe
is in an unknown component, with negative pressure, $p_X \la -\rho_X/3$
(Garnavich \etal, 1998).
The simplest explanation is a cosmological constant with
$\Omega_\Lambda \sim 2/3$.  
(In fact, Equ. 10, which is deeply rooted
in the history of cosmology, is not sufficiently accurate at the
redshifts of the SNe1a being used, and the two groups compute
$d_L \equiv (1+z)r(z)$ as a function of $\Omega_M$ and $\Omega_\Lambda$
and fit to the observations.)

\begin{figure}
\epsfig{file=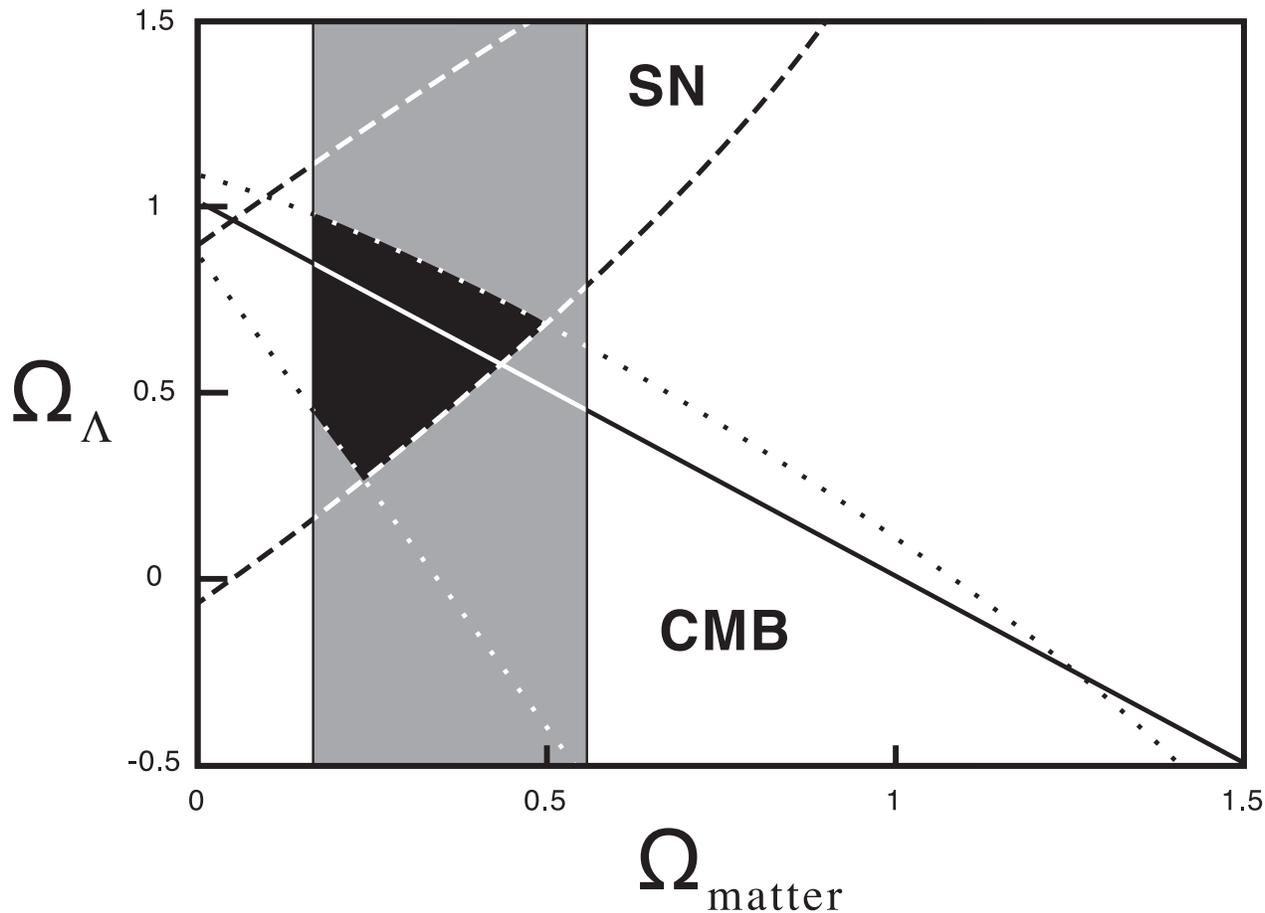,width=5in,angle=90}
\caption{Constraints in the $\Omega_\Lambda$  vs $\Omega_M$ plane.  
Three different types of observations are shown: SNe Ia measures of expansion 
acceleration (SN); the CMB observations of the location of the first acoustic 
peak (CMB); and the determinations of the matter density, $\Omega_M = 0.35
\pm 0.07$ (dark vertical band). The diagonal line indicates a flat Universe,
$\Omega_M + \Omega_{\Lambda} = 1$; regions denote "3-$\sigma$" confidence.
Darkest region denotes the concordance region: $\Omega_\Lambda \sim {2/3}$ and 
$\Omega_M \sim {1/3}$.
}
\label{fig:likelihood}
\end{figure}

\begin{figure}
\epsfig{file=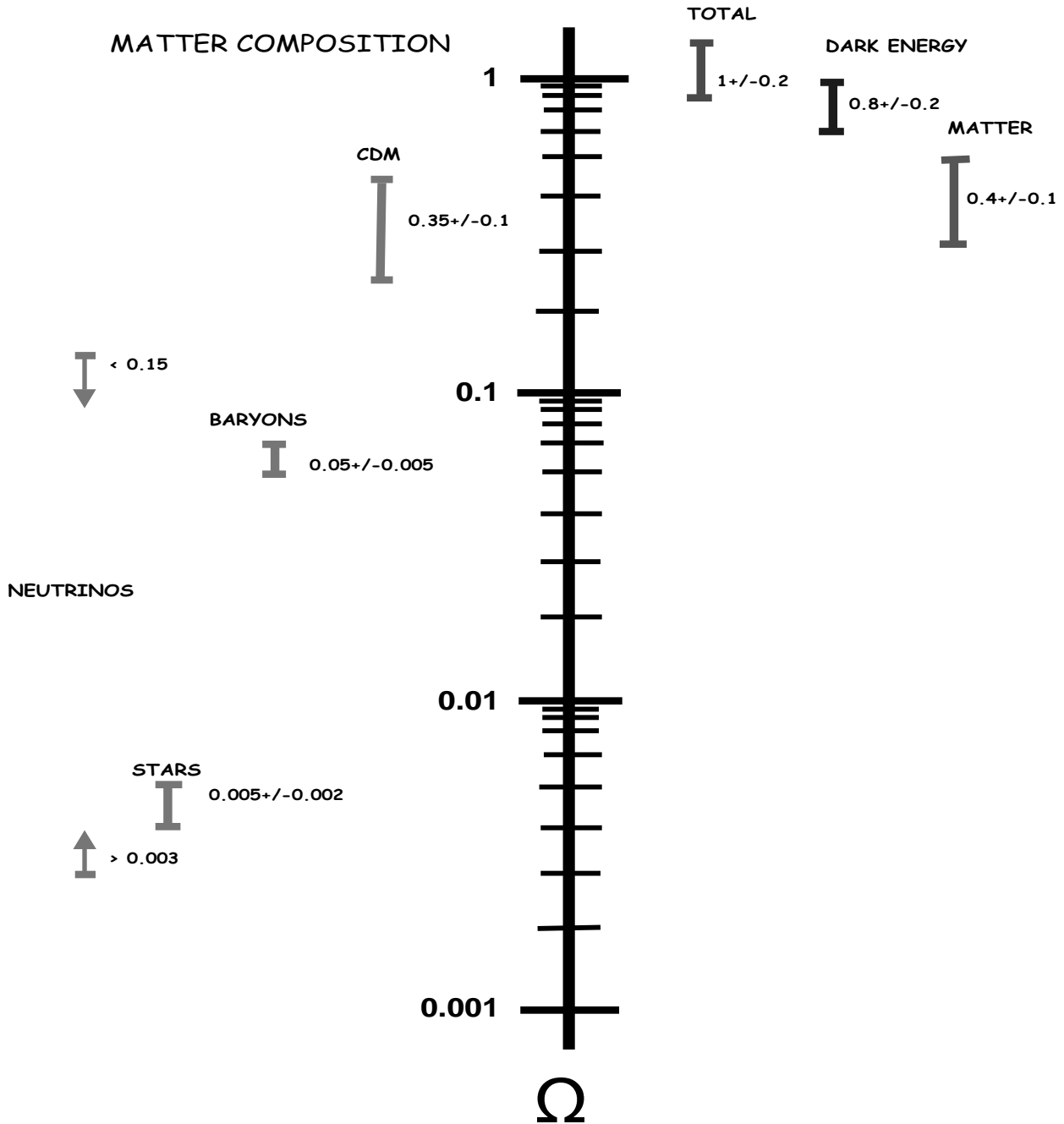,height=7.5in,width=6.5in}
\caption{Summary of matter/energy in the Universe.
The right side refers to an overall accounting of matter
and energy; the left refers to the composition of the matter
component.  The upper limit to mass density contributed
by neutrinos is based upon the failure of the hot dark
matter model and the lower limit follows from the
SuperK evidence for neutrino oscillations.
Dark Energy range is preliminary.
}
\label{fig:Omega-sum}
\end{figure}

Pulling this together, cosmologists for the
first time have a plausible accounting of matter
and energy in the Universe:  stars contribute around 0.4\%
of the critical density, baryons contribute 5\%, nonrelativistic
particles of unknown type contribute 30\%,
and vacuum energy contributes 64\%,
for a total equaling the critical density (see Figures~\ref{fig:likelihood}
and \ref{fig:Omega-sum}).
We should emphasize that plausible does not mean correct.

In addition to the fact that most of the matter and energy
in the Universe is dark, most of the ordinary matter is dark
(i.e., not in bright stars).  The possibilities for the dark baryons
include ``dark stars'' and diffuse hot or warm gas (recall,
in clusters, most of the baryons are in hot, intracluster gas).
Dark stars could take the form of faint, low-mass stars, failed stars
(i.e., objects below the mass required for hydrogen burning,
$M\la 0.08M_\odot$), white dwarfs, neutron stars or black holes.

Most of the mass of our own Milky Way galaxy is dark, existing
in an extended halo (an approximately spherical distribution
of matter with density falling as $1/r^2$).  Unsuccessful
searches for faint stars in our galaxy have eliminated them
as a viable candidate, and theoretical arguments disfavor
white dwarfs, black holes and neutron stars -- all should
lead to the production of more heavy elements than are observed.
Further, the measured rate of star formation indicates that
only a fraction of the baryons have formed into bright, massive stars.

Experimental searches for dark stars in our own galaxy have been
carried out using the gravitational microlensing technique:
dark stars along the line of sight to nearby
galaxies (e.g., the Large and Small Magellanic Clouds
and Andromeda) can gravitationally lens the distant bright
stars, causing a well-defined temporary brightening (Paczynski, 1986).
The results, however, are perplexing (see e.g., Sadoulet, 1999).
More than a dozen such brightenings of LMC stars have been seen, suggesting
that a significant fraction of our galaxy's halo exists
in the form of half-solar mass white dwarfs.  However, such
a population of white dwarfs should be visible,
and they have not been seen.  Because of our imperfect
knowledge of our own galaxy and the LMC, it is possible that
the lenses are not associated with the halo of our galaxy but
rather are low-mass stars in the LMC, in an intervening
dwarf galaxy in between, or are actually in
the disk of our galaxy, if the disk is warped enough to pass
in front of the line to the LMC.

\subsection{Structure formation and primeval inhomogeneity}

The COBE detection of CMB anisotropy on angular scales of
$10^\circ$ was a major milestone (Smoot \etal, 1992), providing the first
evidence for the fluctuations that seeded all the
structure in the Universe and strong evidence for
the gravitational instability picture for structure formation,
as the size of the inhomogeneity was sufficient to explain
the structure observed today.   It also ushered in a powerful
new probe of structure formation and dark matter.  An early
implication of COBE was galvanizing:  nonbaryonic dark matter
is required to explain the structure seen today.  Because
baryons are tightly coupled to photons in the Universe
and thereby supported against gravitational collapse
until after decoupling, larger amplitude density perturbations
are required, which in turn lead to larger CMB temperature
fluctuations than are observed.

Two key issues are the character and origin of the inhomogeneity
and the quantity and composition of matter, discussed above.
It is expected that there is a spectrum of fluctuations, described
by its Fourier decomposition into plane waves.  In addition,
there are two generic types of inhomogeneity:  curvature perturbations,
fluctuations in the local curvature of the Universe which
by the equivalence principle affect all components of the energy
density alike; and isocurvature perturbations, which as their
name indicates are not ingrained in the curvature but arise
as pressure perturbations caused by local changes in
the equation of state of matter and energy in the Universe.

The two most promising ideas for the fundamental origin of the primeval
inhomogeneity are quantum fluctuations which become curvature
fluctuations during inflation (Hawking, 1982; Starobinskii, 1982;
Guth and Pi, 1982; Bardeen, Steinhardt, and Turner, 1983)
and topological defects (such
as cosmic strings) that are produced during a cosmological
phase transition (see e.g., Vilenkin \& Shellard, 1994).  
The inflation scenario will be discussed in
detail later on.  Topological defects produced in a cosmological
symmetry-breaking phase transition around $10^{-36}\sec$ generate
isocurvature fluctuations:  the conversion of energy from radiation
to defects leads to a pressure perturbation that propagates outward
and ultimately leads to a density inhomogeneity.  The defect
scenario is currently disfavored by measurements of CMB anisotropy
(Allen \etal, 1997; Pen \etal, 1997).

One graphic indicator of the progress being made on the large-scale
structure problem is the number of viable models:  the flood of
data has trimmed the field to one or possibly two models.  A few years
ago the defect model was a leading contender; and another, more
phenomenological model put forth by Peebles was also in the running
(Peebles, 1987).
Peebles' model dispensed with nonbaryonic dark matter, assumed
$\Omega_B =\Omega_0 \sim 0.2$, and posited local
variations in the distribution of baryons (isocurvature perturbations)
of unknown origin.  Its demise was CMB anisotropy:  it predicted
too much anisotropy on small angular scales.
The one clearly viable model is cold dark matter plus inflation,
which is discussed below.  The challenge to theorists is make sure that
at least one model remains viable as the quantity and quality
of data improve!

\section{MORE FUNDAMENTAL QUESTIONS}

Beyond the questions involving dark matter and structure
formation, there is a set of more fundamental
questions, ranging from the matter/antimatter asymmetry in
the Universe to the origin of the expansion itself.  For
these questions there are attractive ideas, mainly rooted in the physics of the
early Universe, which remain to be developed, suggesting
that a more fundamental understanding of our Universe is possible.

\smallskip
\noindent{\em Baryon/lepton asymmetry.}  While the laws
of physics are very nearly matter -- antimatter symmetric,
the Universe is not.  On scales as large as clusters of
galaxies there is no evidence for antimatter.  In the context
of the hot big bang, a symmetric Universe would be even more
puzzling:  at early times ($t\ll 10^{-5}\sec$)
matter -- antimatter pairs would
be as abundant as photons, but as the Universe cooled
matter and antimatter would annihilate until nucleons and
antinucleons were too rare to find one another.  This would
result in only trace
amounts of matter and antimatter, a few nucleons and antinucleons
per $10^{18}$ photons, compared to the observed nucleon to
photon ratio:  $\eta \equiv n_N/n_\gamma = (5\pm 0.5)\times 10^{-10}$.

In order to avoid the annihilation catastrophe the early Universe
must possess a slight excess of matter over antimatter, i.e.,
a small net baryon number:  $n_B/n_\gamma \equiv n_b/n_\gamma
-n_{\bar b}/n_\gamma = \eta = 5\times 10^{-10}$.  Such an
initial condition for the Universe seems as odd as having
to assume the $^4$He mass fraction is 25\%.  (Charge
neutrality requires a similar excess of electrons over positrons;
because lepton number can be hidden in the three neutrino species, it is
not possible to say that the total lepton asymmetry is comparable
to the baryon asymmetry.)

A framework for understanding the origin of the baryon asymmetry
of the Universe was put forth in a prescient paper by Sakharov in 1967:
baryon-number violating and matter -- antimatter symmetry violating
interactions occurring in a state of nonequilibrium allow a small,
net baryon number to develop.  If the idea of baryogenesis
is correct, the explanation of the baryon asymmetry is not unlike
that of the primeval $^4$He abundance (produced by nonequilibrium
nuclear reactions).  The key elements of baryogenesis are all in
place:  baryon number is violated in the standard model of particle
physics (by subtle quantum mechanical effects) and in GUTs;
matter -- antimatter symmetry is known to be violated by a small
amount in the neutral Kaon system ($CP$ violation at the level
of $10^{-3}$); and maintaining thermal equilibrium in the expanding
and cooling Universe depends upon whether or not particle interactions
proceed rapidly compared to the expansion rate.  The details
of baryogenesis have not been worked out, and may involve grand
unification physics, but the basic idea is very compelling
(see e.g., Kolb and Turner, 1990).

\smallskip
\noindent{\em The heat of the big bang.}  The entropy associated
with the CMB and three neutrino seas is enormous:  within the
observable Universe, $10^{88}$ in units of $k_B$ (the number of
nucleons is 10 orders of magnitude smaller).  Where did all the
heat come from?  As we discuss in the next section, inflation
may provide the answer.

\smallskip
\noindent{\em Origin of the smoothness and flatness.}  On
large scales today and at very early times the Universe is
very smooth.  (The appearance of inhomogeneity today does
belie a smooth beginning as gravity drives the growth of
fluctuations.)  Since the particle horizon at last-scattering
(when matter and radiation decoupled) corresponds to an angle
of only $1^\circ$ on the sky, the smoothness could not have
arisen via causal physics.  (Within the isotropic and homogeneous
FLRW model no explanation is required of course.)

In a sense emphasized first by Dicke and Peebles (1979)
and later by Guth (1982), the Universe is very flat.  Since $\Omega_0$
is not drastically different from unity, the curvature radius
of the Universe is comparable to the Hubble radius.  During
a matter or radiation dominated phase the curvature radius
decreases relative to the Hubble radius.  This implies that
at earlier times it was even larger than the Hubble radius,
and that $\Omega$ was even closer to one:  $|\Omega - 1| <
10^{-16}$ at 1\,sec.  To arrive at the Universe we see today,
the Universe must have begun very flat (and thus expanding
very close to the critical expansion rate).

The flatness and smoothness problems are not indicative of
any inconsistency of the standard model, but they do require 
special initial conditions.  Stated by
Collins and Hawking (1973), the set of initial conditions
that evolve to a
Universe qualitatively similar to ours is of measure zero.
While not {\it required} by observational data, the inflation 
model addresses both the smoothness and flatness problems.

%% \noindent{\em Very-large scale structure.}

\smallskip
\noindent{\em Origin of the big bang, expansion, and all that.}
In naming the big-bang theory Hoyle tried to call attention to
the colossal big-bang event, which, in the context of general relativity
corresponds to the creation of matter, space and time from a
space-time singularity.  In its success, the big-bang theory is
a theory of the events {\it following} the big-bang singularity.  In
the context of general relativity the big-bang event requires
no further explanation (it is consistent with ``St. Augustine's
principle,'' since time is created along with space, there is
no {\it before} the big bang).  However, many if not most physicists
believe that general relativity, which is a classical theory,
is not applicable any earlier than $10^{-43}\sec$ because
quantum corrections should become very significant, and further,
that a quantum theory of gravity will eliminate the big-bang
singularity allowing the ``before the big-bang question'' to be
addressed.  As we will discuss, inflation addresses the big-bang
question too.

\section{BEYOND THE STANDARD MODEL:  INFLATION + COLD DARK MATTER}

The 1980s were ripe with interesting ideas about the early Universe
inspired by speculations about the unification of the forces and
particles of nature (see e.g., Kolb and Turner, 1990).
For example: relic elementary particles as
the dark matter; topological defects as the seeds for structure
formation; baryon number violation and $C$, $CP$ violation as
the origin of the baryon asymmetry of the Universe (baryogenesis);
and inflation.  From all this, a compelling paradigm for extending
the standard cosmology has evolved:  Inflation + Cold Dark Matter.
It is bold and expansive and is being tested by a flood of observations.
It may even be correct!

The story begins with a brief period of tremendous expansion --
a factor of greater than $10^{27}$ growth in the scale factor in $10^{-32}\sec$.
The precise details of this ``inflationary phase'' are not 
understood, but in most models the exponential expansion is driven by the
(potential) energy of a scalar field initially displaced from the minimum
of its potential energy curve.  Inflation blows up a small, subhorizon-sized 
portion of the Universe to a size much greater than that of
the observable Universe today.  Because this subhorizon-sized
region was causally connected before inflation, it can be expected
to be smooth  -- including the very small portion of it that
is our observable part of the Universe.  Likewise, because
our Hubble volume is but a small part of the region that inflated,
it looks flat, regardless of the initial curvature of the region
that inflated, $R_{\rm curv} \gg H_0^{-1}$, which via the Friedmann
equation implies that $\Omega_0 = 1$.

It is while this scalar field responsible for inflation rolls
slowly down its potential that the exponential expansion takes
place (Linde, 1982; Albrecht \& Steinhardt, 1982).
As the field reaches the minimum of the potential energy
curve, it overshoots and oscillates about it:  the potential
energy of the scalar field has been converted to coherent scalar
field oscillations (equivalently, a condensate of zero momentum
scalar-field particles).  Eventually, these particles decay into
lighter particles which thermalize, thereby explaining the tremendous
heat content of the Universe and ultimately the photons in the CMB
(Albrecht \etal 1982).

Quantum mechanical fluctuations arise in such a scalar field
that drives inflation; they are on truly microscopic scales
($\la 10^{-23}\rcm$).  However, they are stretched in size
by the tremendous expansion during inflation to astrophysical scales.
Because the energy density associated with the scalar field depends upon its
value (through the scalar field potential energy), these fluctuations
also correspond to energy density perturbations, and they are
imprinted upon the Universe as perturbations in the local curvature.
Quantum mechanical fluctuations in the space-time metric give
rise to a stochastic, low-frequency background of gravitational waves.

The equivalence principle holds that local acceleration cannot
be distinguished from gravity; from this it follows that
curvature perturbations ultimately
become density perturbations in all species -- photons, neutrinos,
baryons and particle dark matter.  The shape of the spectrum of perturbations
is nearly scale-invariant.  [Such a form for the spectrum
was first discussed by Harrison (1970). Zel'dovich, who
appreciated the merits of such a spectrum early on, emphasized its
importance for structure formation.]
Scale-invariant refers to the fact that the perturbations in the
gravitational potential have the same amplitude on all length
scales (which is not the same as the density perturbations having the
same amplitude).  When the wavelength of a given mode crosses
inside the horizon ($\lambda = H^{-1}$), the amplitude of the
density perturbation on that scale is equal to the perturbation
in the gravitational potential.

The overall amplitude (or normalization) depends very much upon the
specific model of inflation (of which there are many).  Once the
overall normalization is set, the shape fixes the level of
inhomogeneity on all scales.  The detection of anisotropy on
the scale of $10^\circ$ by COBE in 1992 and the subsequent refinement
of that measurement with the full four-year data set permitted the
accurate (10\%) normalization of the inflationary spectrum of density
perturbations; soon, the term COBE-normalized became
a part of the cosmological vernacular.

On to the cold dark matter part; inflation predicts a flat Universe
(total energy density equal to the critical density).  Since
ordinary matter (baryons) contributes only about 5\%
of the critical density, there must be something else.  The leading
candidate is elementary particles remaining from the earliest moments
of particle democracy.  Generically, they
fall into two classes -- fast moving, or hot dark matter; and
slowly moving, or cold dark matter (see Sadoulet, 1999).
Neutrinos of mass 30\,eV
or so are the prime example of hot dark matter -- they move
quickly because they were once in thermal equilibrium and are
very light.  Axions and neutralinos are examples of cold dark
matter. Neutralinos move slowly because they too were once in
thermal equilibrium and they are very heavy.  Axions are extremely
light but were never in thermal equilibrium (having been produced
very, very cold).

\begin{figure}
\epsfig{file=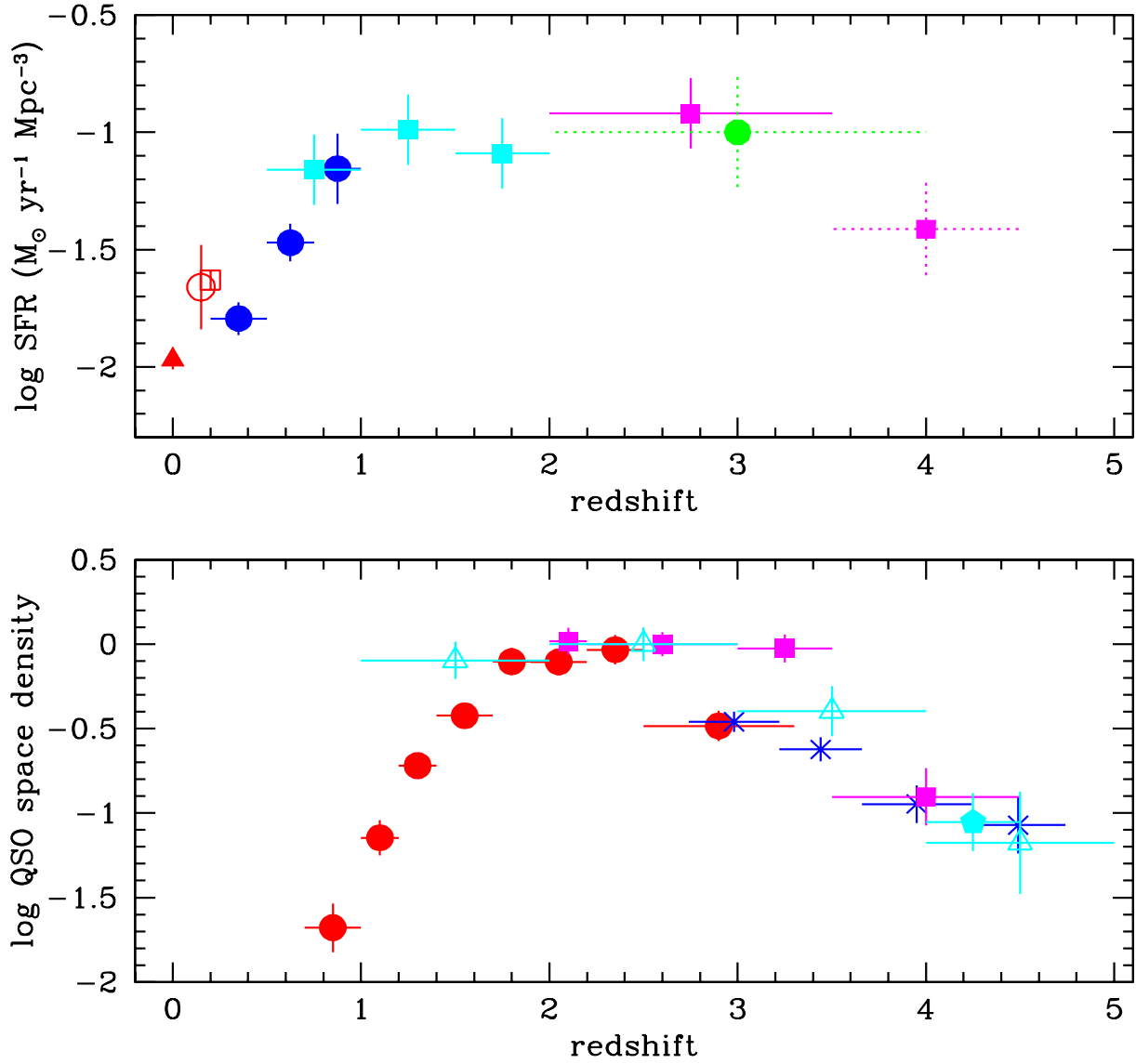,height=6.5in,width=6.7in}
\caption{{\em Top}: Star formation rate in galaxies is plotted vs redshift.
{\em Bottom}: The number density of quasi-stellar objects (galaxies
with accreting black holes) vs redshift.
The points have been corrected for dust and the 
relative space density of QSOs (from Madau, 1999)
}
\label{fig:SFR}
\end{figure}

If most of the matter is hot, then structure in the Universe
forms from the top down:  large things, like superclusters form
first, and fragment into smaller objects such as galaxies.  This
is because fast moving neutrinos smooth out density perturbations
on small scales by moving from regions of high density into
regions of low density (Landau damping or collisionless phase
mixing).  Observations very clearly indicate that galaxies formed
at redshifts $z\sim 2 - 4$ (see Fig. \ref{fig:SFR}), 
before superclusters which are just
forming today. So hot dark matter is out, at least as a major
component of the dark matter (White, Frenk, and Davis 1983).
This leaves cold dark matter.

Cold dark matter particles cannot move far enough to damp perturbations
on small scales, and structure then forms from the bottom up:  galaxies, 
followed by clusters of galaxies, and so on (see e.g., Blumenthal
\etal, 1984).  For COBE-normalized cold
dark matter we can be even more specific.  The bulk of galaxies should
form around redshifts $z\sim 2-4$, just as the observations now indicate.

At present, the cold dark matter + inflation scenario looks very
promising -- it is consistent with a large body of observations:
measurements of the anisotropy of the CMB, redshift surveys of
the distribution of matter today, deep probes of the Universe
(such as the Hubble Deep Field), and
more (see Liddle \& Lyth, 1993 and Fig.~\ref{fig:SFR}). While the evidence
is by no means definitive, and has hardly begun to discriminate
between different inflationary models and versions of CDM,
we can say that the data favor a flat Universe, almost
scale-invariant density perturbations, and cold dark matter with
a small admixture of baryons.

\section{PRECISION COSMOLOGY}

The COBE DMR measurement of CMB anisotropy on the $10^\circ$ angular
scale and determination of the primeval deuterium abundance
served to mark the beginning of a new era of precision cosmology.
Overnight, COBE changed the study of large-scale structure:
for theories like inflation and defects which specify the shape of
the spectrum of density perturbations, the COBE measurement fixed
the level of inhomogeneity on all scales to an accuracy of around
10\%.  Likewise, the measurement of the primeval deuterium abundance,
led to a 10\% determination of the baryon density.

Within the next few years, an avalanche of data, driven by advances
in technology, promises definitive independent observations of the
geometry, mass distribution and composition, and detailed structure
of the Universe. In a radical departure from its history, cosmology
is becoming an exact science.   These new observations span the wavelength
range from microwave to gamma rays and beyond,
and utilize techniques as varied as
CMB microwave interferometry, faint supernova photometry and spectroscopy,
gravitational lensing, and massive
photometric and spectroscopic surveys of millions of galaxies.

The COBE measurement of CMB anisotropy on angular scales
from around $10^\circ$ to $100^\circ$ yielded a precise determination of
the amplitude of mass fluctuations on very large scales, $10^3\Mpc -
10^4\Mpc$.  A host of experiments will view the CMB with
much higher angular resolution and more precision than COBE,
culminating in the two satellite experiments, NASA's MAP and
ESA's Planck Surveyor, which will map the full sky to an
angular resolution of $0.1^\circ$.  In so doing, the mass distribution
in the Universe at a simpler time, before nonlinear structures
had formed, will be determined on scales from $10^4\Mpc$ down to $10\Mpc$.
(Temperature fluctuations on angular scale $\theta$ arise from density
fluctuations on length scales $L\sim 100h^{-1}\Mpc [\theta /{\rm deg}]$;
fluctuations on scales $\sim 1\Mpc$ give rise to galaxies, on scales
$\sim 10\Mpc$ give rise to clusters, and on scales $\sim 100\Mpc$ give
rise to great walls.)

The multipole power spectrum of CMB temperature fluctuations
has a rich structure and encodes a wealth of information about
the Universe.  The peaks in the power spectrum
are caused by baryon-photon oscillations, which are driven
by the gravitational force of the dark matter.  Since
decoupling is essentially instantaneous, different Fourier modes
are caught at different phases, which is reflected in a multipole
spectrum of anisotropy (see Fig.~\ref{fig:CMB-anis}).
The existence of the first peak is evident.
If the satellite missions are as successful as cosmologists
hope, and if foregrounds (e.g., diffuse emission from our own galaxy
and extragalactic point sources) are not a serious problem,
it should be possible to use the measured multipole power
spectrum to determine
$\Omega_0$ and many other cosmological parameters (e.g., $\Omega_Bh^2$,
$h$, $n$, level of gravitational waves, $\Omega_\Lambda$, and
$\Omega_\nu$) to precision of 1\% or better in some cases
(see Wilkinson, 1999). 

Another impressive map is in the works.  The present 3-dimensional
structure of the local Universe will also be mapped to unprecedented precision
in the next few years by the Sloan Digital Sky Survey (SDSS) (see
Gunn \etal, 1998), which will
obtain the redshifts of a million galaxies over 25\% of the northern sky
out to redshift $z\sim 0.1$, and the Two-degree Field Survey
(2dF), which will collect 250,000 redshifts in many $2^\circ$ patches
of the southern sky (Colless, 1988).  
These surveys will cover around 0.1\% of the
observable Universe, and more importantly, will map structure
out to scales of about $500h^{-1}\Mpc$, well beyond the size of
the largest structures known.  This should be large enough 
to provide a typical sample of the Universe.  The two maps --
CMB snapshot of the Universe at 300,000\,yrs and the SDSS map
of the Universe today -- when used together have enormous leverage
to test cosmological models and determine cosmological parameters.

Several projects are underway to map smaller, more distant parts
of the Universe to study the ``recent'' evolution of galaxies and
structure.  Using a new large spectrograph, the 10-meter Keck
telescope will begin to map galaxies in
smaller fields on the sky out to redshifts of $4$ or so.  Ultimately,
the Next Generation Space Telescope, which is likely to
have an 8-meter mirror and capability in the infrared (most of
the light of high-redshift galaxies has been shifted into
the infrared) will probe the first generation of stars and galaxies.

Much of our current understanding of the Universe is based on the
assumption that light traces mass, because telescopes detect light
and not mass.  There is some evidence that light is not a terribly
``biased'' tracer of mass, at least on the scales of galaxies. However,
it would be a convenient accident if the mass-to-light ratio
were universal.  It is possible that there is a lot of undiscovered
matter, perhaps even enough to bring $\Omega_M$ to unity, 
associated with dim galaxies or other mass concentrations that are 
not correlated with bright galaxies.

Gravitational lensing is a powerful means of measuring
cosmic mass overdensities in the linear regime directly (see e.g.,
Blandford \& Narayan, 1992; Tyson, 1993):
dark matter overdensities at moderate redshift ($z \sim 0.2 - 0.5$)
systematically distort background galaxy images (referred to
as weak gravitational lensing).
A typical random one-square-degree patch of the sky contains
a million faint high-redshift galaxies.  Using these galaxies,
weak gravitational lensing may be used to
map the intervening dark matter overdensities directly.
This technique has been used to map
known or suspected mass concentrations in clusters of galaxies
over redshifts $0.1 < z < 0.8$ (see Fig.~\ref{fig:0024} and Clowe,
\etal, 1998) and only recently has been applied to random fields.
Large mosaics of CCDs make this kind of direct
mass survey possible, and results from these surveys
are expected in the coming years.

\begin{figure}
\epsfig{file=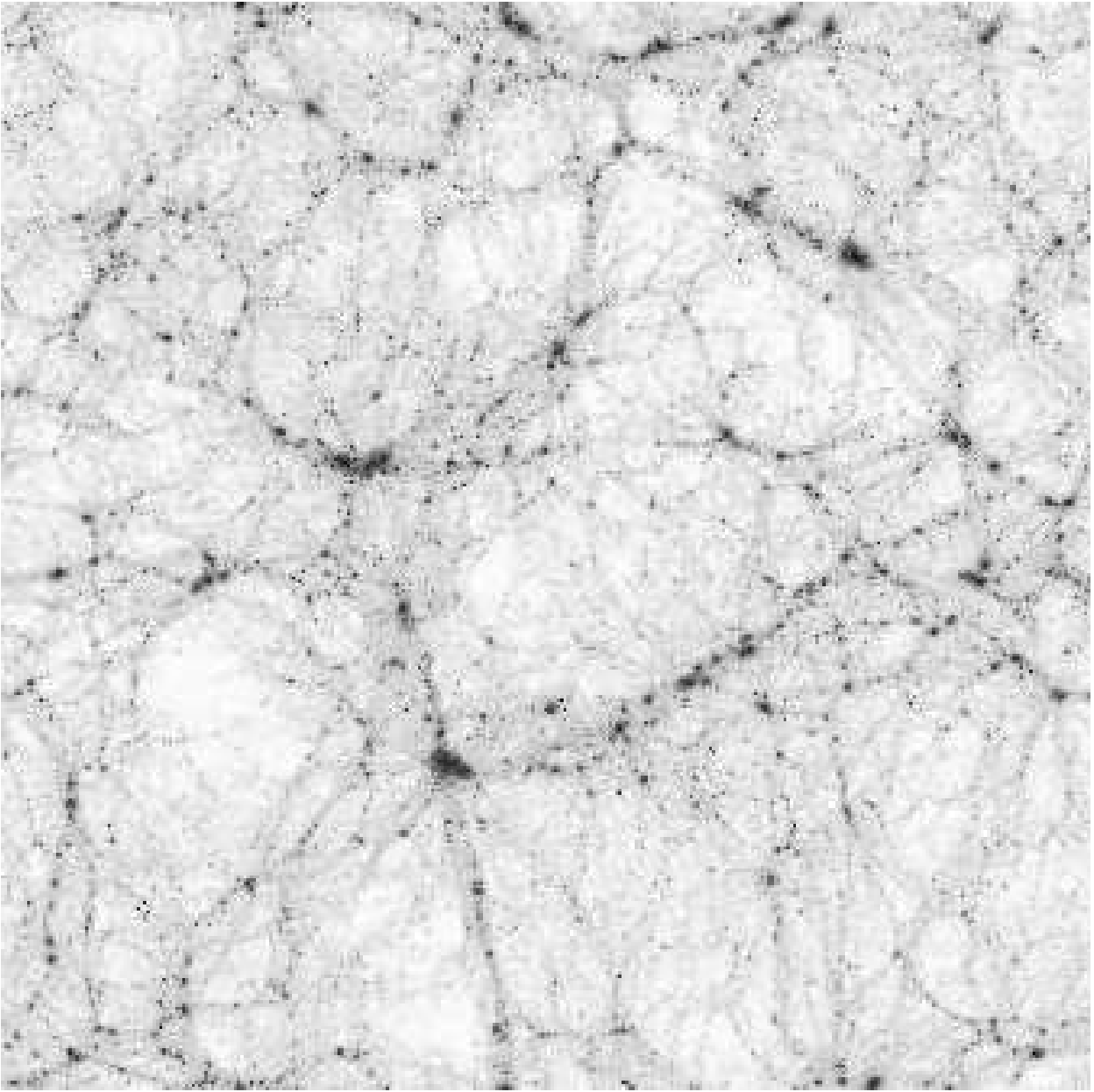,height=6.5in,width=6.5in}
\caption{One of the largest simulations of the development of
structure in the Universe. (from Virgo
Collaboration, 1998). Shown here is projected mass in a $\Lambda CDM$ 
simulation with $256^3$ particles and $\Omega_M = 0.3$, 240 h$^{-1}$ $\Mpc$ 
on a side. The map shown in Fig. \protect\ref{fig:0024} would correspond to a 
window 0.5 $\Mpc$ across, centered on one of the minor mass concentrations.
}
\label{fig:Virgo}
\end{figure}

Crucial to taking advantage of the advances in our understanding
of the distribution of matter in the Universe and the formation
of galaxies are the numerical simulations that link theory with
observation.  Simulations now involve billions of particles,
allowing a dynamical range of a factor of one thousand (see
Fig.~\ref{fig:Virgo}).  Many simulations now involve not only
gravity, but the hydrodynamics
of the baryons.  Advances in computing have been crucial.

Impressive progress has been made toward measuring the cosmological
parameters $H_0$, $q_0$ and $t_0$, and more progress is on the horizon.
A 5\% or better measurement of the
Hubble constant on scales that are a substantial fraction
of the distance across the Universe may be within our grasp.
Techniques that do not rely upon phenomenological standard candles
are beginning to play an important role.
The time delay of a flare event seen in the multiple images of
a lensed quasar is related only
to the redshifts of the lens and quasar, the lens magnification,
the angular separation of the quasar images, and the Hubble constant.
Thanks to a recent flare, an accurate time delay between the two images of the
gravitationally lensed quasar Q0957+561 has been reliably determined,
but the lens itself must be mapped before $H_0$ is precisely determined 
(Kundic \etal, 1997).  This technique is being applied to other
lensed quasar systems as well.  The pattern
of CMB anisotropy has great potential to accurately determine $H_0$.
Another technique (Sunyaev-Zel'dovich, or SZ), which uses the
small distortion of the CMB when viewed through a cluster containing
hot gas (due to Compton up-scattering of CMB photons), has begun to
produce reliable numbers (Birkinshaw, 1998).

Currently, the largest gap in our knowledge of the mass content of
the Universe is identifying the bulk of the matter density, $\Omega_?
=\Omega_M - \Omega_B \sim 0.3$.  The most compelling idea is that
this component consists of relic elementary particles, such as
neutralinos, axions or neutrinos.  If such particles compose most
of the dark matter, then they should account for most of the
dark matter in the halo of our own galaxy and have a local mass
density of around $10^{-24}\gcmm3$.  Several laboratory experiments are
currently running with sufficient sensitivity to search directly
for neutralinos of mass $10\GeV - 500\GeV$ and cross-section that is
motivated by the minimal supersymmetric standard model.  While the
supersymmetric parameter space spans more than 3 orders-of-magnitude
in cross section, even greater sensitivities are expected in the
near future.  These experiments involve high-sensitivity, low-background
detectors designed to detect the small (order keV) recoil energy
when a neutralino elastically scatters off a nucleus in the detector;
the small rates (less than one scattering per day per kg of detector)
add to the challenge (Sadoulet, 1999).

An axion detector has achieved sufficient
sensitivity to detect halo axions, and is searching the mass range
$10^{-6}\eV - 10^{-5}\eV$ where axions would contribute significantly
to the mass density.  This detector, based upon the conversion of
axions to photons in strong magnetic field, consists of a hi-Q
cavity immersed in a 7 Tesla magnetic field and is operating with
a sensitivity of $10^{-23}\,$W in the GHz frequency range.  Within
five years it is hoped that the entire theoretically favored mass
range will be explored (Rosenberg, 1998).

While light neutrinos are no longer favored by cosmologists for
the dark matter, as they would lead to structure in the Universe
that is not consistent with what we see today, because of their
large numbers, $113\cmm3$, they could be an important component
of mass density even if only one species has a tiny mass:
\begin{equation}
\Omega_\nu = \sum_i (m_{\nu_i} /90h^2\eV) \qquad
\Omega_\nu / \Omega_{\rm lum} \simeq \sum_i (m_{\nu_i}/0.2\eV ) \qquad
\Omega_\nu / \Omega_B \simeq \sum_i (m_{\nu_i}/2\eV )
\end{equation}
Even with a mass as small as one 
eV neutrinos would make an imprint on the structure of the Universe
that is potentially detectable.

Particle theorists strongly favor the idea that neutrinos
have small, but nonzero mass, and the see-saw mechanism can
explain why their masses are so much smaller than the
other quarks and leptons:  $m_\nu \sim m_{q,l}^2/M$ where
$M \sim 10^{10}\GeV - 10^{15}\GeV$ is the very large mass
of the right-handed partner(s) of the usual left-handed
neutrinos (see e.g., Schwarz \& Seiberg, 1999).
Because neutrino masses are a fundamental prediction
of unified field theories, much effort is directed at probing
neutrino masses.  The majority of experiments now involve
looking for the oscillation of one neutrino species into
another, which is only possible if neutrinos have mass.
These experiments are carried out at accelerators, at nuclear
reactors, and in large-underground detectors such as Super-Kamiokande
and the SNO facility.

Super-K detects neutrinos from the sun
and those produced in the earth's atmosphere by cosmic-ray interactions.
For several years now the solar-neutrino data has shown evidence
for neutrino oscillations, corresponding to a neutrino mass-difference
squared of around $10^{-5}\eV^2$ or $10^{-10}\eV^2$, too small
to be of cosmological interest (unless two neutrino species are
nearly degenerate in mass.)  The Super-K collaboration recently
announced evidence for neutrino oscillations based upon the atmospheric
neutrino data.  Their results, which indicate a mass-difference squared of
around $10^{-3}-10^{-2}\eV^2$
(Fukuda \etal, 1998) and imply at least one
neutrino has a mass of order $0.1\eV$ or larger, are
much more interesting cosmologically.  Over the next decade
particle physicists will pursue neutrino mass with a host of new
experiments, characterized by very long baselines (neutrino source
and detector separated by hundreds of kilometers) and should clarify
the situation.

\subsection{Testing Inflation + CDM in the precision era}

As we look forward to the abundance (avalanche!) of high-quality observations
that will test Inflation + CDM, we have to make sure the predictions
of the theory match the precision of the data.  In
so doing, CDM + Inflation becomes a theory with ten or more parameters.
For cosmologists,
this is a bit daunting, as it may seem that a ten-parameter
theory can be made to fit any set of observations.  This will
not be the case when one has the quality and quantity of data
that are coming.  The standard model of particle physics offers
an excellent example:  it is
a 19-parameter theory, and because of the high quality of
data from experiments at high-energy accelerators and
other facilities, it has been rigorously tested,
with parameters measured to a precision of better than 1\%
in some cases.

In fact, the ten parameters of CDM + Inflation
are an opportunity rather than a curse:  Because the parameters
depend upon the underlying inflationary model and fundamental
aspects of the Universe, we have the very real possibility of learning
much about the Universe, inflation, and perhaps fundamental
physics.  The ten parameters can be split into two groups:
cosmological and dark matter.

\vspace{10pt}
\centerline{\it Cosmological Parameters}

\begin{enumerate}

\item $h$, the Hubble constant in units of $100\kms\Mpc^{-1}$.

\item $\Omega_Bh^2$, the baryon density.

\item $n$, the power-law index of the scalar density perturbations.
CMB measurements indicate $n=1.1\pm 0.2$; $n=1$ corresponds to
scale-invariant density perturbations.  Several popular
inflationary models predict $n\simeq 0.95$; range of predictions
runs from $0.7$ to $1.2$.

\item $dn/d\ln k$, ``running'' of the scalar index with comoving scale
($k=$ wavenumber).  Inflationary models predict a value of
${\cal O}(\pm 10^{-3})$ or smaller.

\item $S$, the overall amplitude squared of density perturbations,
quantified by their contribution to the variance of the
quadrupole CMB anisotropy.

\item $T$, the overall amplitude squared of gravitational waves,
quantified by their contribution to the variance of the
quadrupole CMB anisotropy.  Note, the COBE normalization determines
$T+S$ (see below).

\item $n_T$, the power-law index of the gravitational wave spectrum.
Scale invariance corresponds to $n_T=0$; for inflation, $n_T$
is given by $-T/7S$. 

\end{enumerate}

\smallskip
\centerline{\it Dark-matter Parameters}
\vspace{10pt}
\begin{enumerate}

\item $\Omega_\nu$, the fraction of critical density in neutrinos
($=\sum_i m_{\nu_{\it i}}/90h^2$).  While the hot dark matter
theory of structure
formation is not viable, it is possible that a small fraction of
the matter density exists in the form of neutrinos.

\item $\Omega_X$, the fraction of critical density in a smooth component
of unknown composition and negative pressure ($w_X \la -0.3$); the
simplest example is a cosmological constant ($w_X = -1$).

\item $g_*$, the quantity that counts the number of ultra-relativistic
degrees of freedom (at late times).  The standard cosmology/standard
model of particle physics predicts $g_* = 3.3626$ (photons in the
CMB + 3 massless neutrino species with temperature $(4/11)^{1/3}$
times that of the photons).  The amount of radiation controls when
the Universe becomes matter-dominated and thus affects the present
spectrum of density fluctuations.

\end{enumerate}

The parameters involving density and gravitational-wave
perturbations depend directly upon the inflationary potential.
In particular, they can be expressed in terms of the potential
and its first two derivatives (see e.g., Lidsey \etal, 1997):

\begin{eqnarray}
S  & \equiv & {5\langle |a_{2m}|^2\rangle \over 4\pi} \simeq
         2.2\,{V_*/m_{\rm Pl}^4 \over (m_{\rm Pl} V_*^\prime /V_*)^2}\\
n -1 & = & -{1\over 8\pi}\left({\mpl V^\prime_* \over V_*} \right)^2
        + {\mpl \over 4\pi}\left( {\mpl V_*^{\prime}\over V_*} \right)^\prime \\
T  & \equiv & {5\langle |a_{2m}|^2\rangle \over 4\pi} =
        0.61 (V_*/m_{\rm Pl}^4)
\end{eqnarray}
where $V(\phi )$ is the inflationary potential, prime denotes $d/d\phi$,
and $V_*$ is the value of the scalar potential when the present horizon
scale crossed outside the horizon during inflation.

As particle physicists can testify, testing a ten (or more) parameter
theory is a long, but potentially rewarding process.  To begin, one has
to test the basic tenets and consistency of the underlying theory.
Only then, can one proceed to take
full advantage of the data to precisely measure parameters of the theory.
The importance of establishing a theoretical framework is illustrated
by measurements of the number of light neutrino species derived from
the decay width of the $Z^0$ boson:  $N_\nu = 3.07\pm 0.12$ (not assuming
the correctness of the standard model); $N_\nu = 2.994\pm 0.012$
(assuming the correctness of the standard model).

In the present case, the putative theoretical framework
is Inflation + CDM, and its basic tenets are: a flat, critical
density Universe;
a nearly scale-invariant spectrum of Gaussian density perturbations;
and a stochastic background of gravitational waves.  The first two
predictions are much more amenable to testing, by a combination of
CMB anisotropy and large-scale structure measurements.  For example,
a flat Universe with Gaussian curvature perturbations implies a multipole
power spectrum of well defined acoustic peaks, beginning at $l\simeq
200$ (see Fig.~\ref{fig:CMB-anis}).  In addition, there are consistency tests:
comparison of the precise BBN determination of the baryon density with
that derived from CMB anisotropy; an accounting of the dark matter and
dark energy by gravitational lensing; SNe1a measurements of acceleration;
and comparison of the different determinations of the Hubble constant.
Once the correctness and consistency of Inflation + CDM has been verified --
assuming it is -- one can zero in on the remaining parameters (subset 
of the list above) and hope to determine them with precision.

\subsection{Present status of Inflation + CDM}

A useful way to organize the different CDM models is by their
dark matter content; within each CDM family, the cosmological
parameters can still vary:  sCDM (for simple), only CDM and baryons;
$\tau$CDM:  in addition to CDM and baryons additional
radiation (e.g., produced by the decay of an unstable
massive tau neutrino); $\nu$CDM: CDM, baryons, and a dash of hot
dark matter (e.g., $\Omega_\nu = 0.2$); and $\Lambda$CDM:
CDM, baryons, and a cosmological constant (e.g., $\Omega_\Lambda = 0.6$).
In all these models, the total energy density sums to the critical
energy density; in all but $\Lambda$CDM, $\Omega_M =1$.

\begin{figure}
\epsfig{file=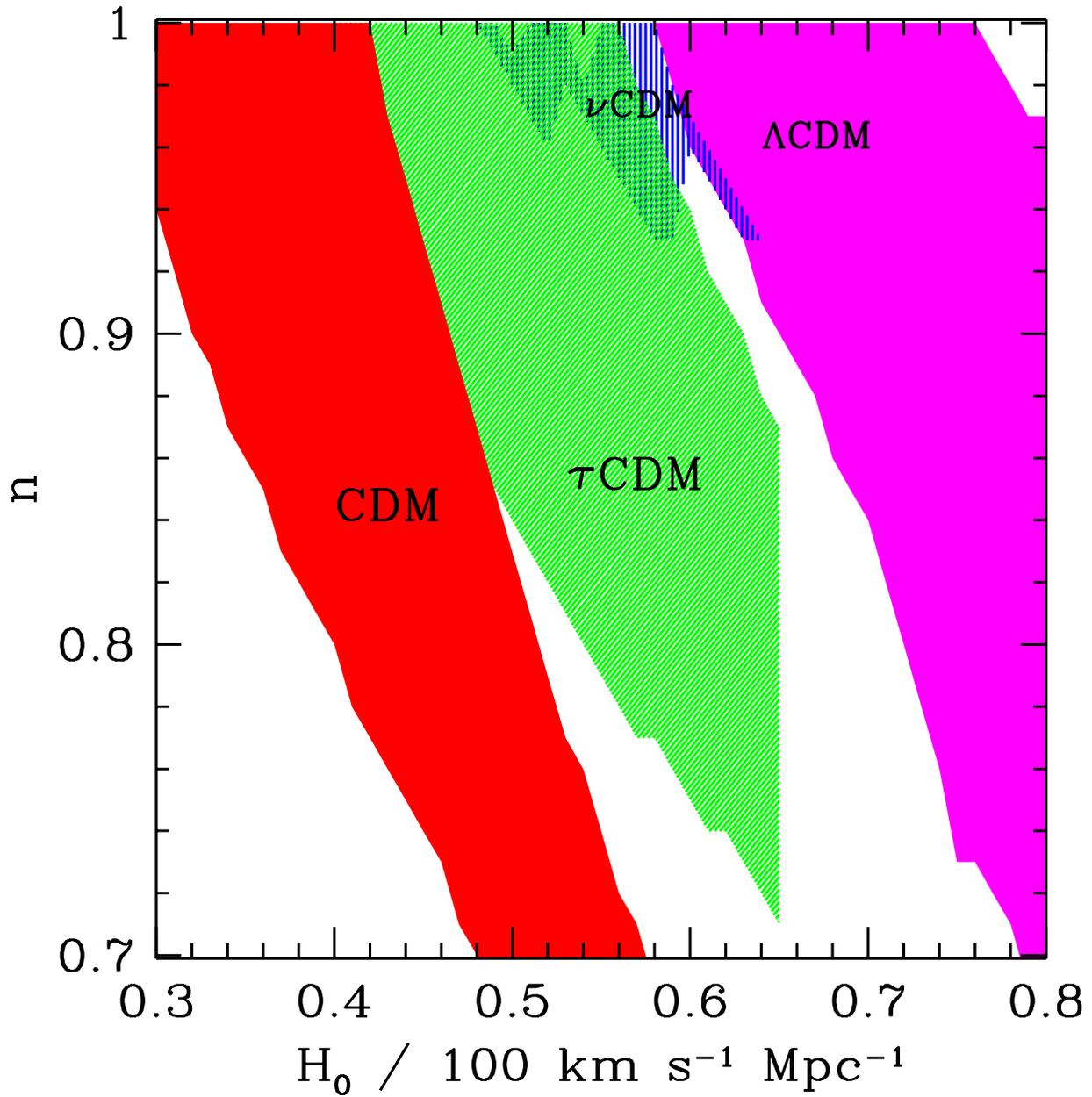,height=6.5in,width=6.4in}
\caption{Acceptable cosmological parameters for different
CDM models, as are characterized by their invisible
matter content: simple CDM (CDM), CDM plus cosmological constant
($\Lambda$CDM), CDM plus some hot dark matter ($\nu$CDM), and CDM plus 
added relativistic particles ($\tau$CDM) (from Dodelson \etal, 1996).
}
\label{fig:cdm-sum}
\end{figure}

\begin{figure}
\epsfig{file=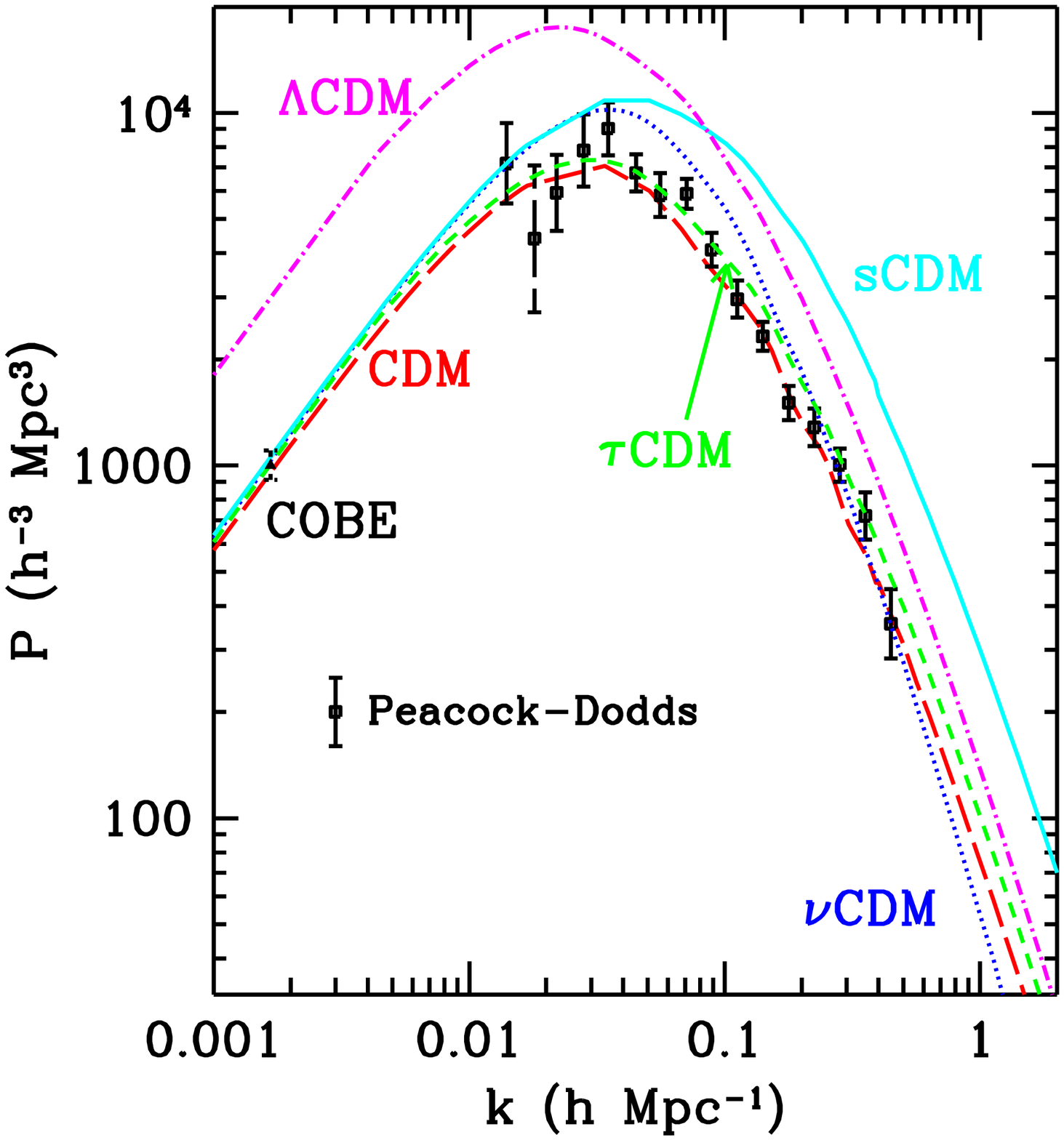,height=6.5in,width=6.4in}
\caption{The power spectrum of fluctuations today, as traced by
bright galaxies (light), as derived from redshift surveys assuming
light traces mass (Peacock and Dodds, 1994).  The curves correspond to the 
predictions of various cold dark matter models.   The relationship
between the power spectrum and CMB anisotropy
in a $\Lambda$CDM model is different, and in fact, the $\Lambda$CDM
model shown is COBE normalized.
}
\label{fig:pd}
\end{figure}

Figure \ref{fig:cdm-sum} summarizes the viability of these different
CDM models, based upon CMB measurements and current determinations of
the present power spectrum of fluctuations (derived from
redshift surveys; see Fig.~\ref{fig:pd}).
sCDM is only viable for low values of the
Hubble constant (less than $55\kms\Mpc^{-1}$) and/or
significant tilt (deviation from scale invariance); the region
of viability for $\tau$CDM is similar to sCDM, but shifted
to larger values of the Hubble constant (as large as
$65\kms\Mpc^{-1}$).  $\nu$CDM has an island of viability
around $H_0\simeq 60\kms\Mpc^{-1}$ and $n\simeq 0.95$.  $\Lambda$CDM
can tolerate the largest values of the Hubble constant.

Considering other relevant data too -- e.g.,
age of the Universe, determinations of $\Omega_M$,
measurements of the Hubble constant, and limits to
$\Omega_\Lambda$ -- $\Lambda$CDM emerges as the `best-fit CDM model'
(see e.g., Krauss \& Turner, 1995; Ostriker \& Steinhardt, 1995).
Moreover, its `key signature,' $q_0\sim -0.5$, may have been confirmed.
Given the possible systematic uncertainties
in the SNe1a data and other measurements, it is premature to conclude that
$\Lambda$CDM is anything but the model to take aim at!

\section{THE NEXT HUNDRED YEARS}

The progress in cosmology over the last hundred years has
been stunning.  With the hot big-bang cosmology we
can trace the history of the Universe to within
a fraction of a second of the beginning.  
Beyond the standard cosmology, we have promising ideas,
rooted in fundamental theory,
about how to extend our understanding to even earlier
times addressing more profound questions, e.g., inflation
+ cold dark matter.  While it remains to be seen whether
or not the expansion is accelerating,
it is a fact that our knowledge of the Universe is accelerating,
driven by new observational results.
Cosmology seems to be in the midst of a golden age;
within ten years we may have a cosmological theory that explains
almost all the fundamental features of the Universe,
the smoothness and flatness, the heat of the CMB, the
baryon asymmetry, and the origin of structure.

There are still larger questions to be answered and to be
asked.  What is the global topology of the Universe?
Did the Universe begin with more than four dimensions?
Is inflation the dynamite of the big bang, and were there
other such big bangs?  Are there cosmological signatures of the
quantum gravity epoch?  

It is difficult -- and dangerous -- to speculate where
cosmology will go in the next twenty years, let alone
the next hundred.  One can never predict the
serendipitous discovery that radically transforms our understanding.
In an age of expensive, complex, and highly focused experiments,
we must be especially vigilant and keep an open mind.  And it
can be argued that the two most important discoveries in cosmology
-- the expansion and the CMB -- were unexpected.
In astrophysics, it is usually a safe bet that things are
more complicated than expected.  But then again, Einstein's
ansatz of large-scale homogeneity and isotropy -- made to make
the equations of general relativity tractable -- turned out to
be a remarkably good description of the Universe.

{}

\clearpage

\clearpage

\clearpage

\clearpage

\clearpage

\clearpage

\clearpage

\clearpage

\clearpage

\clearpage

\clearpage

\clearpage

\clearpage

\clearpage

\begin{thebibliography}{}

\bibitem[]{} Albrecht, A. \& P.J.~Steinhardt, 1982, Phys. Rev. Lett. 48, 1220.

\bibitem[]{} Albrecht, A. \etal, 1982, Phys. Rev. Lett. 48, 1437.

\bibitem[]{} Allen, B. \etal, 1997, Phys. Rev. Lett. 79, 2624.

\bibitem[]{} Bahcall, N. \etal, 1993, Astrophys. J. 415, L17.

\bibitem[]{} Bardeen, J., P. J. Steinhardt, and M. S. Turner, 1983, Phys. Rev. D 28, 679.

\bibitem[]{} Baum, W. A., 1957, AJ 62, 6.

\bibitem[]{} Birkinshaw, M., 1998, Phys. Rep., in press (astro-ph/9808050).

\bibitem[]{} Blandford, R. D. \& R. Narayan, 1992, Ann. Rev.
Astron. Astrophys. 30, 311.

\bibitem[]{} Blumenthal, G., S. Faber, J. Primack, and M. Rees,
1984, Nature 311, 517.

\bibitem[]{} Broadhurst, T.J., R. Ellis, D.C. Koo, and A.S. Szalay, 1990,
Nature 343, 726.

\bibitem[]{} Burles, S. and D. Tytler, 1998a, Astrophys. J. 499, 699.

\bibitem[]{} Burles, S. and D. Tytler, 1998b, Astrophys. J., 507, 732.

\bibitem[]{} Carlberg, R. G., H. K. C. Lee, E. Ellingson, R. Abraham, P. Gravel,
S. Morris, and C. J. Pritchet, 1996, Astrophys. J. 462, 32.

\bibitem[]{} Carlberg, R. G., H. K. C. Lee, E. Ellingson, 
1997, Astrophys. J. 478, 462.

\bibitem[]{} Carlstrom, J., 1999, Physica Scripta, in press.

\bibitem[]{} Carroll, S. M., W. H. Press, and E. L. Turner, 1992,
Ann. Rev. Astron. Astrophys. 30, 499.

\bibitem[]{} Chaboyer, B., P. Demarque, P. Kernan, and L. Krauss, 1998,
Astrophys. J. 494, 96.

\bibitem[]{} Clowe, D., G.A. Luppino, N. Kaiser, J.P. Henry, and
I.M. Gioia, 1998, Astrophys. J. 497, L61.

\bibitem[]{} Colless, M. 1998, Phil. Trans. R.Soc. Lond. A, in press;
astro-ph/9804079.

\bibitem[]{} Collins, C.B. \& and S.W.~Hawking, 1973,
Astrophys. J. 180, 317.

\bibitem[]{} Cowan, J., F.~Thieleman, and J.~Truran, 1991,
Ann. Rev. Astron. Astrophys. 29, 447.

\bibitem[]{} Dekel, A., D. Burstein, and S.D.M. White, 1997,
in {\it Critical Dialogues in Cosmology}, ed. N. Turok (World
Scientific, Singapore).

\bibitem[]{} Dicke, R.H. \& Peebles, P.J.E., 1979, in {\it General Relativity:
An Einstein Centenary Survey}, edited by S. Hawking and W. Israel (Cambridge
Univ. Press, Cambridge), p. 504.

\bibitem[]{} Dodelson, S., E.I. Gates, and M.S. Turner, 1996,
Science 274, 69.

\bibitem[]{} Evrard, A.E., 1997, MNRAS 292, 289.

\bibitem[]{} Filippenko A.V., and A.G. Riess, 1998 (astro-ph/9807008).

\bibitem[]{} Fixsen, D. J., \etal, 1996, Astrophys. J. 473, 576.

\bibitem[]{} Fukuda, Y. \etal (SuperKamiokande Collaboration), 1998,
Phys. Rev. Lett. 81, 1562.

\bibitem[]{} Gaillard, M.K., P. Grannis, and F. Sciulli, 1999,
Rev. Mod. Phys., in press.

\bibitem[]{} Garnavich, P., \etal, 1998, Astrophys. J.,
in press (astro-ph/9806396).

\bibitem[]{} Ge, J., J. Bechtold, and J.H. Black, 1997, Astrophys. J.
474, 67.

\bibitem[]{} Geller, M., and J. Huchra, 1989, Science 246, 897.

\bibitem[]{} Gunn, J.E., M. Carr, C. Rockosi, and M. Sekiguchi, 1998, Astron. J.,
in press.

\bibitem[]{} Guth, A., 1982, Phys. Rev. D 23, 347.

\bibitem[]{} Guth, A. and S.-Y. Pi, 1982, Phys. Rev. Lett. 49, 1110.

\bibitem[]{} Harrison, E.R., 1970,  Phys. Rev. D 1, 2726.

\bibitem[]{} Hawking, S.W. 1982, Phys. Lett. B 115, 295.

\bibitem[]{} Jungman, G., M. Kamionkowski, and K. Griest, 1996,
Phys. Rep. 267, 195.

\bibitem[]{} Jungman, G., M. Kamionkowski, A. Kosowsky, and
D.N. Spergel, 1996, Phys. Rev. Lett. 76, 1007.

\bibitem[]{} Kolb, E. W., and M. S. Turner, 1990, {\it The Early Universe} (Addison-
Wesley, Redwood City).

\bibitem[]{} Kundic, T. \etal, 1997, Astrophys. J. 482, 75.

\bibitem[]{} Krauss, L. \& M.S. Turner, 1995, Gen. Rel. Grav. 27, 1137.

\bibitem[]{} Leibundgut, B. \etal, 1996, Astrophys. J. Lett. 466, 21.

%%\bibitem {} Levin, J. J., etal., 1997, Phys. Rev. Lett. 79, 974.

\bibitem[]{} Liddle, A. \& Lyth, 1993, D. Phys. Rep. 231, 1.

\bibitem[]{} Lidsey, J. \etal, 1997, Rev. Mod. Phys. 69, 373.

\bibitem[]{} Linde, A., 1982, Phys. Lett. B 108, 389.

\bibitem[]{} Madau, P. 1999, Physics Scripta.

\bibitem[]{} Madore, B. \etal, 1009, Nature 395, 47.

\bibitem[]{} Mather, J.C., \etal, 1990, Astrophys. J. 354, L37.

\bibitem[]{} Oort, J.H., 1983, Ann. Rev. Astron. Astrophys. 21, 373.

\bibitem[]{} Ostriker, J. P., and P. J. Steinhardt, 1995, Nature 377, 600.

\bibitem[]{} Oswalt, T. D., J. A. Smith, M. A. Wood, and P. Hintzen, 1996,
Nature 382, 692.

\bibitem[]{} Paczynski, B., 1986, Astrophys. J. 304, 1.

\bibitem[]{} Peacock, J. \& S.~Dodds, 1994, MNRAS 267, 1020.

\bibitem[]{} Peebles, P.J.E., 1987, Nature 327, 210.

\bibitem[]{} Peebles, P. J. E., D. N. Schramm, E. L. Turner and R. G. Kron,
1991, Nature 352, 769.

\bibitem[]{} Peebles, P. J. E., 1993, {\it Principles of Physical Cosmology}
(Princeton University Press).

\bibitem[]{} Pen, U.-L., \etal, 1997, Phys. Rev. Lett. 79, 1611.

\bibitem {} Perlmutter, S., \etal, 1998, Nature 391, 51.

\bibitem[]{} Riess, A. G., W. H. Press, and R. P. Kirshner, 1996,
Astrophys. J. 473, 88.

\bibitem[]{} Riess, A. G. \etal, 1998, Astron. J. 116, 1009.

\bibitem[]{} Rosenberg, L. J., 1998, PNAS 95, 59.

\bibitem[]{} Sadoulet, B., 1999, Rev. Mod. Phys., in press.

\bibitem[]{} Sandage, A. 1961, ApJ 133, 355.

\bibitem[]{} Schmidt, B.P. \etal 1998, Astrphys. J. 507, 46.

\bibitem[]{} Schramm, D. and M. Turner, 1998, Rev. Mod. Phys. 70, 303.

\bibitem[]{} Schwarz, J. \& N. Seiberg, 1999, Rev. Mod. Phys.

\bibitem[]{} Shectman, S. \etal, 1996, Astrophys. J. 470, 172.

\bibitem[]{} Sloan Digital Sky Survey (SDSS); see http://www.sdss.org.

\bibitem[]{} Smoot, G. \etal, 1992, Astrophys. J. 396, L1.

\bibitem[]{} Songaila, A., L.L. Cowie, M. Keane, A.M. Wolfe, E.M. Hu,
A.L. Oren, D. Tytler, and K.M. Lanzetta, 1994, Nature 371, 43.

\bibitem[]{} Starobinskii, A.A., 1982, Phys. Lett. B 117, 175.

\bibitem[]{} Steigman, G., D.N. Schramm, and J. Gunn, 1977,
Phys. Lett. B 66, 202.

\bibitem[]{} Two-degree Field (2dF); see http://msoww.anu.edu.au/$\sim$colless/2dF/

\bibitem[]{} Tyson, J. A., 1993, Phys. Today 45, 24.

\bibitem[]{} Tyson, J.A., G. Kochanski, and I.P. Dell'Antonio, 1998,
Astrophys. J. 498, L107.

\bibitem[]{} Vilenkin, A. \& E.P.S. Shellard, 1994, {\em Cosmic
Strings and other Topological Defects} (Cambridge Univ. Press,
Cambridge).

\bibitem[]{} Virgo Collaboration; 
see http://www.mpa-garching.mpg.de/$\sim$jgc/sim\_virgo.html

\bibitem[]{} Weinberg, S., 1972, {\it Gravitation and  Cosmology:
Principals and Applications of the General Theory of Relativity}
(Wiley, New York).

\bibitem[]{} Weinberg, S. 1989, Rev. Mod. Phys. 61, 1.

\bibitem[]{} White, S.D.M., C. Frenk, and M. Davis, 1983,
Astrophys. J. 274, L1.

\bibitem[]{} Wilkinson, D.T., 1999, Rev. Mod. Phys., in press.

\bibitem[]{} Willmer, C.N.A., D.C. Koo, A.S. Szalay, and M.J. Kurtz,
1994, Astrophys. J. 437, 560.


\end{thebibliography}
\end{document}